\documentclass[nofootinbib,aps,a4paper,letterpaper,superscriptaddress,
twocolumn,showpacs,times,eqsecnum]{revtex4}
\usepackage{amsmath}
\usepackage{amsfonts}
\usepackage{booktabs}
\usepackage{multirow}
\usepackage{siunitx} 
\usepackage{adjustbox}
\pdfoutput=1
\usepackage{graphicx}
\usepackage{color}
\usepackage{braket}
\usepackage{dcolumn}
\usepackage{bm,url}
\usepackage[linktocpage]{hyperref}
\usepackage{subfigure}
\usepackage{amsfonts}
\usepackage[usenames,dvipsnames,svgnames]{xcolor}  
\usepackage{hyperref}   
\definecolor{oxfordblue}{rgb}{0.0, 0.13, 0.28}
\definecolor{burgundy}{rgb}{0.5, 0.0, 0.13}
\definecolor{darkolivegreen}{rgb}{0.33, 0.42, 0.18}
\definecolor{darkblue}{rgb}{0,0,0.5}
\definecolor{richcarmine}{rgb}{0.84, 0.0, 0.25}
\definecolor{darkblue}{rgb}{0,0,0.5}
\definecolor{bluer}{rgb}{0.00,0.50,0.75}{}
\hypersetup{colorlinks=true, citecolor=red, linkcolor=blue,
 urlcolor = magenta, filecolor=magenta}

\begin{document}
 
 \newcommand\be{\begin{equation}}
  \newcommand\ee{\end{equation}}
 \newcommand\bea{\begin{eqnarray}}
  \newcommand\eea{\end{eqnarray}}
 \newcommand\bseq{\begin{subequations}} 
  \newcommand\eseq{\end{subequations}}
 \newcommand\bcas{\begin{cases}}
  \newcommand\ecas{\end{cases}}
 \newcommand{\p}{\partial}
 \newcommand{\f}{\frac}

 \title{From thermogravity to thermocosmology with a running non-conservation scale}

 \author{Emmanuel N. Saridakis}
 \email{msaridak@noa.gr}
 \affiliation{Institute for Astronomy, Astrophysics, Space Applications and
Remote Sensing, National Observatory of Athens, 15236 Penteli, Greece}
 \affiliation{Departamento de Matem\'{a}ticas, Universidad Cat\'{o}lica del 
  Norte, Avda. Angamos 0610, Casilla 1280, Antofagasta, Chile}
 \affiliation{CAS Key Laboratory for Research in Galaxies and Cosmology, 
School 
  of Astronomy and Space Science,
  University of Science and Technology of China, Hefei 230026, China}

\begin{abstract}
Thermogravity offers a novel approach to cosmic acceleration,
deriving trace-free gravitational equations and matter
non-conservation from thermodynamic processes in spacetime.
We extend this framework towards thermocosmology by introducing
a running non-conservation scale $L_2(H)$, while restricting
non-conservation to cold dark matter. We investigate two
prescriptions, namely a Hubble-proportional rate and an
inverse-Hubble running model, deriving their cosmological
dynamics and effective geometric dark-energy description.
The first admits an exact analytical solution and is related
to quadratic running-vacuum cosmologies, while the second
approaches standard matter dilution at early times and
allows reversals in the direction of energy exchange.
Interestingly, the reconstructed geometric dark energy
can exhibit quintessence-like and phantom-like behavior,
including smooth crossings of the phantom divide, without
introducing a phantom field or superacceleration.
A unified dynamical-system analysis reveals distinct
matter-era critical configurations and a continuous family
of de Sitter solutions that are transversely attracting
under suitable conditions.
Confronting both models with Pantheon+ Type Ia
supernovae, cosmic chronometers, and DESI DR2
baryon acoustic oscillations, we find a mild
preference for positive present-day matter
non-conservation. The Akaike information criterion
slightly favors the thermogravity scenarios,
whereas the Bayesian information criterion favors
the simpler $\Lambda$CDM model. Thus,   running thermocosmology provides
a viable framework for nontrivial effective
dark-energy dynamics.
\end{abstract}

 \pacs{04.50.Kd, 98.80.-k, 95.36.+x, 98.80.Es}
 \maketitle

 \section{Introduction}

Despite the remarkable success of the standard $\Lambda$CDM
cosmological model in describing the evolution of the Universe,
the nature of its dark components remains unknown. In
particular, the physical origin of dark matter and dark energy,
together with the small observed value of the cosmological
constant compared to theoretical expectations, constitute
fundamental open problems in modern cosmology
\cite{Weinberg:1988cp,Bertone:2004pz}.
Furthermore, the increasing precision of cosmological
observations has revealed possible tensions between different
measurements, while recent results from the Dark Energy
Spectroscopic Instrument (DESI) have renewed   interest in
a possible evolution of dark energy
\cite{Abdalla:2022yfr,CosmoVerseNetwork:2025alb}.
These issues have motivated several extensions of the standard
cosmological paradigm. One possibility is to retain general
relativity as the underlying gravitational theory and introduce
additional particles, fields, or effective fluids, such as
quintessence, k-essence, or interacting dark-energy models
\cite{Copeland:2006wr,Cai:2009zp}. Alternatively, one may modify the
gravitational sector itself
\cite{CANTATA:2021asi,Clifton:2011jh,Capozziello:2011et,Nojiri:2017ncd}, leading to a
variety of theories
that can provide an effective description of dark energy
through gravitational dynamics. Representative examples
include $f(R)$ gravity  \cite{Starobinsky:1980te, Capozziello:2002rd,
DeFelice:2010aj} and  $f(G)$ gravity \cite{Nojiri:2005jg, DeFelice:2008wz}, based on
curvature,
 $f(T)$ gravity
\cite{Cai:2015emx, Ferraro:2006jd,
  Linder:2010py, Chen:2010va,Tamanini:2012hg},  $f(T, T_G)$ gravity
 \cite{Kofinas:2014owa, Kofinas:2014daa}, and  $f(T, B)$ gravity
 \cite{Bahamonde:2015zma, Bahamonde:2016grb},
based on torsion, and  $f(Q)$ theories
\cite{BeltranJimenez:2017tkd,BeltranJimenez:2019tme,Anagnostopoulos:2021ydo,
Heisenberg:2023lru} and $f(Q,C)$ gravity
\cite{De:2023xua}, based on non-metricity.

A conceptually different approach arises from the relation
between gravitation and thermodynamics
\cite{Jacobson:1995ab,Padmanabhan:2003gd,Padmanabhan:2009vy}. The
discovery of
black-hole entropy and temperature  \cite{Bekenstein:1973ur,Hawking:1975vcx}
 suggested that gravitational
systems possess genuine thermodynamic properties, leading to
the possibility that gravitational dynamics may emerge from
the collective behavior of microscopic spacetime degrees
of freedom. In particular,
Einstein's equations can be obtained as thermodynamic
equations of state by applying the Clausius relation to
local causal horizons \cite{Frolov:2002va,Cai:2005ra,Akbar:2006kj,Eling:2006aw}.
This interpretation has subsequently been developed in
several directions, establishing a connection between
gravitational field equations, horizon thermodynamics,
and the microscopic structure of spacetime  \cite{Lymperis:2018iuz,
Moradpour:2016rcy,Moradpour:2017ycq,
Moradpour:2018ivi,MohseniSadjadi:2010nu, Bamba:2018zil,
 Hernandez-Almada:2021rjs,
 Asghari:2021bqa,  Nojiri:2021jxf,
Giardino:2020myz, Barrow:2020kug, Leon:2021wyx,
 Jizba:2022bfz,
Odintsov:2022qnn,  Dheepika:2022sio,
Nojiri:2022nmu,
Lambiase:2023ryq, Mohammadi:2023lss, Salehi:2023gzv,
Jizba:2023fkp,    Jizba:2024klq,
 Mondal:2024wno, Asghari:2024sbu,
Karabat:2024trf,Luciano:2026ufu,Leizerovich:2026pfy,Luciano:2026eiy,
Leizerovich:2026qie} (for a review see
\cite{Saridakis:2026vxo}).
An interesting question within this framework is whether
the usual gravitational equations exhaust the possible
thermodynamic description of spacetime, or whether more
general thermodynamic processes may give rise to different
macroscopic gravitational dynamics.

Recently, the thermodynamic interpretation of gravity was extended by considering an
infinitesimal Otto-like cycle within a local causal diamond, including work-producing
processes beyond conventional heat exchange \cite{Isichei:2025ssf}.
The resulting framework, termed thermogravity, leads to trace-free gravitational
field equations, a preferred thermodynamic frame, and modified
energy-momentum conservation. While vacuum-energy contributions cancel from the
trace-free equations, matter non-conservation can induce a dynamical
geometrical contribution, affecting cosmological evolution.
The cosmological implications of the basic scenario have been investigated  in
\cite{Lee:2026ayb,Brereton:2026mki}.
However, the basic scenario  assumes a constant characteristic
non-conservation scale $L_2$, and thus its possible evolution
with the cosmological background remains unexplored. On the other hand, allowing
this scale to vary may   lead to
additional cosmological dynamics without modifying
the trace-free gravitational equations.

The possibility that cosmological quantities evolve with
the expansion has been extensively investigated in the
literature. In particular, Hubble-dependent vacuum energy
and running cosmological parameters have been considered
in dynamical-vacuum and renormalization-group approaches
\cite{Shapiro:2000dz,Basilakos:2009wi,
Lima:2013dmf,Basilakos:2013vya,
Sola:2016jky,SolaPeracaula:2023swx,
Bonanno:2001xi,Reuter:2003ca}, or in  generalized equations of state
\cite{Nojiri:2005sr,Nojiri:2006zh,Nojiri:2009kx,Nojiri:2017opc}.
Similarly, expansion-dependent matter-production rates
and dynamical dark-sector interactions have been
investigated in various cosmological frameworks
\cite{Prigogine:1988jax,Lima:1995xz,
Basilakos:2010yp,Lima:2012cm,
Wang:2015wga,Yang:2019uzo,Yang:2022csz,
Silva:2025hxw}.
These developments provide further motivation for
exploring the cosmological implications of an evolving
thermodynamic non-conservation scale.

In the present work, we extend thermogravity towards
thermocosmology by introducing a running non-conservation
scale $L_2(H)$, where $H$ is the Hubble function.
We focus on the realization in which only cold dark matter
undergoes non-conservation, while baryons and radiation
retain their standard evolution. We consider two specific
prescriptions for the corresponding rate
$\Gamma(H)=L_2^{-1}(H)$, namely a Hubble-proportional model
and a more general model involving a linear dependence
on the inverse Hubble function. These choices allow
us to investigate different non-conservation histories,
including cases where the relative non-conservation
rate is suppressed at early times or changes sign
during cosmological evolution. We derive the corresponding
background equations and examine the evolution of the
reconstructed geometric dark-energy contribution,
which can exhibit effective quintessence-like
or phantom-like behavior, as well as a crossing of
the phantom divide. We additionally perform a
dynamical-system analysis, identifying the relevant
cosmological critical points and the conditions for
late-time de Sitter attraction. Finally, we confront
both models with observations from Pantheon+ Type Ia
supernovae, cosmic chronometers, and DESI DR2 baryon
acoustic oscillations. As we show, both prescriptions
provide a good description of the late-time observations,
with modest improvements in the goodness of fit relative
to $\Lambda$CDM. The Akaike information criterion
slightly favors the thermogravity scenarios, whereas
the Bayesian information criterion continues to favor
the simpler $\Lambda$CDM model.

The manuscript is organized as follows. In Sec.~\ref{sec:thermogravity}, we review
the basic thermogravity framework and introduce the running
non-conservation scale, deriving the corresponding
cosmological equations and the effective geometric
dark-energy description. In
Sec.~\ref{sec:cosmological_dynamics}, we investigate
the two specific models, examining their analytical
and numerical background evolution and their
dynamical-system properties. In
Sec.~\ref{sec:observational_constraints}, we present
the observational constraints and compare their
statistical performance with $\Lambda$CDM and
constant-rate thermogravity. Finally, in Sec. \ref{sec:conclusions}
we summarize our results and discuss the
corresponding conclusions.

\section{Thermogravity with a running non-conservation scale}
\label{sec:thermogravity}

In this section, we present the theoretical framework
of thermogravity and introduce its extension to a running
non-conservation scale. We first review the gravitational
field equations and the associated matter non-conservation
law, and then derive the cosmological equations for a
Hubble-dependent scale. Finally, we examine the resulting
effective geometric dark-energy contribution and its
relation to matter non-conservation.

\subsection{Gravitational field equations and non-conservation law}
\label{subsec:field_equations}

The thermodynamic interpretation of gravity provides a framework
in which gravitational field equations emerge as macroscopic
relations associated with the thermodynamic properties of
spacetime \cite{Jacobson:1995ab,Padmanabhan:2009vy}.
In particular, the application of the Clausius relation to
local causal horizons allows Einstein's equations to be
recovered as thermodynamic equations of state.
Recently, Isichei and Magueijo \cite{Isichei:2025ssf}
extended this approach by considering a non-degenerate
thermodynamic cycle that includes work-producing processes
in addition to heat exchange.
The resulting formulation of thermogravity allows for
controlled departures from local Lorentz invariance and
energy-momentum conservation, opening the possibility
of cosmological dynamics beyond those of general relativity.

The underlying construction involves an infinitesimal
Otto-like thermodynamic cycle within a local causal diamond.
Besides the conventional entropy $S$ and temperature $T$,
an additional thermodynamic variable ${\mathcal{N}}$ is introduced,
representing an effective number of microscopic spacetime
constituents, together with its conjugate chemical potential
$\mu$. The variation of the internal energy is then written as
$
dU=T\,dS+\mu\,d{\mathcal{N}}$.
In the degenerate limit, the additional work contribution
vanishes and the standard thermodynamic description of
gravity is recovered. Including the work-producing processes
leads to a nonzero cycle efficiency, parametrized by
\begin{equation}
\eta=\frac{2}{15}\frac{\ell}{L_2},
\label{eq:thermo_efficiency}
\end{equation}
where $\ell$ is the characteristic size of the causal
diamond and $L_2$ is a new length scale controlling the
additional thermodynamic effects. The linear dependence
on $\ell$ is imposed in Ref.~\cite{Isichei:2025ssf} to
obtain well-defined macroscopic field equations in the
infinitesimal-diamond limit. Moreover, the conventional description
is recovered as $L_2^{-1}\rightarrow0$.

The gravitational field equations retain the trace-free form
\begin{equation}
R_{\mu\nu}-\frac{1}{4}R g_{\mu\nu}
=
8\pi G
\left(
T_{\mu\nu}-\frac{1}{4}T g_{\mu\nu}
\right),
\label{eq:thermo_field_equations}
\end{equation}
where $R_{\mu\nu}$ and $R$ are the Ricci tensor and scalar,
respectively, while $T_{\mu\nu}$ is the total
energy-momentum tensor and $T=g^{\mu\nu}T_{\mu\nu}$ its trace.
Throughout the manuscript, we work in units where $c=1$.
Since contributions proportional to the metric cancel
identically, a constant vacuum-energy term does not enter
Eq.~\eqref{eq:thermo_field_equations}. However, the trace-free
equations do not independently determine the trace of the
gravitational field equations. Taking their covariant
divergence and using the contracted Bianchi identity yields
\begin{equation}
\nabla_\mu T^\mu{}_\nu
=
\frac{1}{32\pi G}
\nabla_\nu\left(R+8\pi G T\right).
\label{eq:thermo_bianchi_identity}
\end{equation}
For a conserved energy-momentum tensor, the quantity
$R+8\pi G T$ is constant, recovering the usual integration
freedom associated with the cosmological constant. However, when matter conservation
is relaxed, this quantity
may instead evolve with the cosmological background.

The additional thermodynamic relation introduces a
preferred frame, characterized by a timelike unit vector
$n^\mu$ satisfying $n_\mu n^\mu=-1$.
Defining the spatial projection tensor
$h_{\mu\nu}=g_{\mu\nu}+n_\mu n_\nu$, the non-conservation
law   takes the form \cite{Isichei:2025ssf}
\begin{equation}
-n_\mu\nabla_\nu T^{\mu\nu}
=
\frac{1}{L_2}
\left(
n^\mu n^\nu+\frac{1}{3}h^{\mu\nu}
\right)T_{\mu\nu}.
\label{eq:thermo_nonconservation}
\end{equation}
The contraction on the right-hand side corresponds to
a light-cone average of the energy-momentum tensor in
the preferred frame. For a perfect fluid comoving with
this frame, with
$T_{\mu\nu}=(\rho+p)n_\mu n_\nu+p g_{\mu\nu}$,
it reduces to $\rho+p$. Consequently, in  a homogeneous and isotropic
Friedmann-Lemaître-Robertson-Walker (FLRW)  cosmological background
\begin{equation}
ds^2=-dt^2+a^2(t)
\left[
\frac{dr^2}{1-Kr^2}+r^2d\Omega^2
\right],
\label{eq:FLRW_metric}
\end{equation}
where $K$ denotes the spatial curvature parameter,
    relation (\ref{eq:thermo_nonconservation})
gives
\begin{equation}
\dot{\rho}+3H(\rho+p)
=
\frac{\rho+p}{L_2},
\label{eq:thermo_fluid_nonconservation}
\end{equation}
with $H\equiv\dot{a}/a$   the Hubble function.
A positive source term corresponds to energy production,
whereas a negative one describes energy depletion.
In the cosmological generalization considered below,
we allow the effective inverse scale $L_2^{-1}$ to take
either sign. Negative values should be understood as a
phenomenological extension of the original thermodynamic
construction, rather than as a microscopic derivation
of a negative cycle efficiency.

For a cosmological medium containing several matter
species, the non-conservation law can be generalized to
\begin{equation}
-n_\mu\nabla_\nu T_I^{\mu\nu}
=
\left(
n^\mu n^\nu+\frac{1}{3}h^{\mu\nu}
\right)
\sum_J M_{IJ}T_{\mu\nu}^{J},
\label{eq:thermo_multicomponent}
\end{equation}
where $I,J$ label the different components and
$M_{IJ}$ specifies the distribution of non-conservation
effects among them. The universal choice
$M_{IJ}=\delta_{IJ}/L_2$ introduces non-conservation
in all matter components. However, such a prescription
would also modify the standard evolution of baryons
and radiation, with possible implications for
cosmological observables.

In the present work, we adopt the non-universal realization
considered in \cite{Lee:2026ayb}, in which
non-conservation is restricted to cold dark matter.
The coupling matrix is therefore specified by
\begin{equation}
M_{cc}=\frac{1}{L_2},
\qquad
M_{IJ}=0
\quad \text{for} \quad (I,J)\neq(c,c),
\label{eq:thermo_cdm_matrix}
\end{equation}
where $c$ denotes cold dark matter.
Accordingly, baryons, photons, and neutrinos satisfy
their standard conservation equations, while cold dark
matter receives the additional thermodynamic source term.
This choice is phenomenological rather than a unique
prediction of the microscopic construction. Nevertheless,
it allows us to investigate the cosmological consequences
of thermogravity while preserving the conventional
background evolution of baryons and radiation.

Having specified the gravitational equations and the
matter non-conservation law, we now extend the framework
by allowing the effective non-conservation scale to
vary with the cosmological expansion.

 \subsection{Running non-conservation scale and cosmological equations}
\label{subsec:running_scale}

In the original thermogravity construction, the non-conservation
scale $L_2$ is constant \cite{Isichei:2025ssf}.
However, its possible dependence on the cosmological background
has not been determined by the underlying microscopic description. In particular,
the original thermogravity construction
allows for temperature-dependent characteristic scales
$L_i(T)$ when the assumed thermodynamic decoupling is
relaxed \cite{Isichei:2025ssf}. This provides additional
motivation for investigating a cosmologically evolving
non-conservation scale, although its specific dependence
on $H$ remains phenomenological.
In the present work, we consider the phenomenological
generalization
\begin{equation}
L_2=L_2(H),
\label{eq:running_L2}
\end{equation}
where $H$ is the Hubble parameter.
This prescription introduces a time-dependent strength of
matter non-conservation while preserving the trace-free
gravitational field equations.
We emphasize that the dependence on $H$ is assumed at the
cosmological background level, rather than derived from
the microscopic thermodynamic construction. Specifically,
the Hubble parameter describes the expansion of the
cosmological congruence and should not be identified
directly with a local thermodynamic temperature.

We introduce the non-conservation rate
\begin{equation}
\Gamma(H)\equiv\frac{1}{L_2(H)},
\label{eq:Gamma_definition}
\end{equation}
which has dimensions of inverse time and determines
the strength of the additional contribution to the
matter continuity equation. The original constant-scale
thermogravity model is recovered for constant $\Gamma$,
whereas $\Gamma=0$ corresponds to standard matter
conservation. Within our phenomenological extension,
positive and negative values of $\Gamma$ describe
energy production and depletion, respectively.
A running rate can also change sign during cosmological
evolution. In this case, $\Gamma(H)$ is the convenient
quantity to parametrize, since it can pass smoothly
through zero even though the corresponding effective
length scale becomes unbounded.

For the cold-dark-matter-only realization introduced
in the previous subsection, the non-conservation matrix
has the single non-vanishing component
$M_{cc}=\Gamma(H)$. The continuity equations become
\begin{align}
\dot{\rho}_c+3H\rho_c
&=\Gamma(H)\rho_c,
\label{eq:running_rhoc}\\
\dot{\rho}_b+3H\rho_b
&=0,
\label{eq:running_rhob}\\
\dot{\rho}_r+4H\rho_r
&=0,
\label{eq:running_rhor}
\end{align}
where the subscripts $c$, $b$, and $r$ denote cold
dark matter, baryons, and radiation, respectively.
Thus, baryons and radiation retain their standard
evolution, with $\rho_b=\rho_{b0}a^{-3}$ and
$\rho_r=\rho_{r0}a^{-4}$, while the cold dark matter
density is modified by the non-conservation source.
Here and in the following, we normalize the present
scale factor to $a_0=1$, and we use the subscript ``0'' for the value of a quantity
at present.

For the case of  FLRW spacetime given in (\ref{eq:FLRW_metric}), the preferred
thermodynamic frame is identified with
the cosmological rest frame, $n^\mu=(1,0,0,0)$,
for which $\nabla_\mu n^\mu=3H$.
Substituting the above metric into the trace-free
gravitational equations (\ref{eq:thermo_field_equations}),
we obtain
\begin{equation}
\dot H-\frac{K}{a^2}
=
-4\pi G\sum_I(\rho_I+p_I),
\label{eq:running_Hdot_general}
\end{equation}
where the sum extends over the physical matter
components. The non-conservation rate does not
appear explicitly in this equation, but affects
the gravitational dynamics through the evolution
of the matter densities.

In the following, we restrict ourselves to a spatially
flat Universe, $K=0$, containing cold dark matter,
baryons, and radiation. Equation
(\ref{eq:running_Hdot_general}) then reduces to
\begin{equation}
\dot H
=
-4\pi G
\left(
\rho_c+\rho_b+\frac{4}{3}\rho_r
\right).
\label{eq:running_Hdot}
\end{equation}
Together with Eqs.~(\ref{eq:running_rhoc})-(\ref{eq:running_rhor}), this relation
determines
the homogeneous cosmological evolution for a
specified non-conservation rate $\Gamma(H)$.

To express the equations in dimensionless form,
we introduce the number of e-folds $N\equiv\ln a$
and the normalized Hubble parameter
\begin{equation}
E(N)\equiv\frac{H(N)}{H_0},
\label{eq:running_E_definition}
\end{equation}
where $H_0$ is the present Hubble parameter.
We further define the present density parameters
$\Omega_{I0}=8\pi G\rho_{I0}/(3H_0^2)$ and the
normalized cold dark matter density
\begin{equation}
f_c(N)\equiv\frac{\rho_c(N)}{\rho_{c0}}.
\label{eq:running_fc_definition}
\end{equation}
Finally, we introduce
\begin{equation}
\xi(E)\equiv
\frac{\Gamma(H_0E)}{H_0E}
=
\frac{\Gamma(H)}{H},
\label{eq:running_xi_definition}
\end{equation}
which measures the non-conservation rate relative
to the cosmological expansion. Its present value
is $\xi_0=\xi(1)=\Gamma(H_0)/H_0$.

The cosmological equations can then be written
as the coupled system
\begin{align}
\frac{dE^2}{dN}
&=
-3\Omega_{b0}e^{-3N}
-4\Omega_{r0}e^{-4N}
-3\Omega_{c0}f_c(N),
\label{eq:running_E2}\\
\frac{df_c}{dN}
&=
\left[-3+\xi(E)\right]f_c(N),
\label{eq:running_fc}
\end{align}
where $\Omega_{c0}=\Omega_{m0}-\Omega_{b0}$. Additionally,
the present-time normalization conditions are
$
E(0)=1$ and $f_c(0)=1$.
For a specified function $\Gamma(H)$, these equations
form a closed system for the cosmological background.

The modified cold dark matter continuity equation
can also be integrated formally, yielding
\begin{equation}
f_c(N)
=
e^{-3N}
\exp\left[
\int_0^N
\xi\big(E(\tilde N)\big)\,d\tilde N
\right].
\label{eq:running_fc_integral}
\end{equation}
Thus, deviations from the standard dust scaling
depend on the integrated non-conservation history.
For a positive $\Gamma(H)$ throughout the evolution,
the cold dark matter density at a given past redshift
is smaller than in the conserved case, when the same
present density is imposed. A negative rate produces
the opposite effect, while a sign-changing rate can
lead to different contributions over successive
cosmological epochs.

For the subsequent analysis, it is convenient to
express the equations in terms of redshift,
$1+z=a^{-1}$. Equations (\ref{eq:running_E2})
and (\ref{eq:running_fc}) become
\begin{align}
\frac{dE^2}{dz}
&=
3\Omega_{b0}(1+z)^2
+4\Omega_{r0}(1+z)^3
+\frac{3\Omega_{c0}f_c(z)}{1+z},
\label{eq:running_E2_redshift}\\
\frac{df_c}{dz}
&=
\frac{3-\xi(E)}{1+z}f_c(z).
\label{eq:running_fc_redshift}
\end{align}
These equations will be used to determine the
background evolution of the two specific models
considered below and their corresponding
cosmological observables.

Finally, we mention that although the trace-free gravitational
equations determine the evolution of the Hubble
parameter, they do not independently impose an
algebraic Friedmann constraint. The corresponding
integration contribution, and its interpretation
in terms of an effective geometric dark-energy
sector, will be examined in the next subsection.

\subsection{Matter non-conservation and effective geometric dark energy}
\label{subsec:effective_geometric}

The trace-free gravitational equations, together with the
modified matter conservation law, admit an Einstein-like
reconstruction involving an effective geometrical contribution
\cite{Isichei:2025ssf}. In particular, the freedom associated
with the missing trace allows this contribution to evolve
when the matter energy-momentum tensor is not conserved.
As we show below, the resulting cosmological equations
exhibit formal similarities with interacting dark-energy
models \cite{Amendola:1999er,Wang:2016lxa,Wang:2024vmw}, although their physical
origin is different.

We begin by rewriting
Eq.~\eqref{eq:thermo_field_equations} in the equivalent form
\begin{equation}
G_{\mu\nu}+\Lambda_{\mathrm{eff}}g_{\mu\nu}
=
8\pi G T_{\mu\nu},
\label{eq:effective_Einstein}
\end{equation}
where $G_{\mu\nu}=R_{\mu\nu}-Rg_{\mu\nu}/2$ is the Einstein
tensor, and
\begin{equation}
\Lambda_{\mathrm{eff}}
\equiv
\frac{1}{4}\left(R+8\pi G T\right).
\label{eq:Lambda_eff_definition}
\end{equation}
Here, $T_{\mu\nu}$ contains the physical matter and radiation
components, without an additional dark-energy fluid.
The quantity $\Lambda_{\mathrm{eff}}$ represents the
geometrical contribution associated with the missing trace
of the gravitational field equations, rather than a
fundamental cosmological constant or an independently
introduced vacuum-energy source.

Taking the covariant divergence of
Eq.~\eqref{eq:effective_Einstein} and using the contracted
Bianchi identity, we obtain
\begin{equation}
\nabla_\nu\Lambda_{\mathrm{eff}}
=
8\pi G\nabla^\mu T_{\mu\nu}.
\label{eq:Lambda_eff_divergence}
\end{equation}
Thus, when the matter energy-momentum tensor is conserved,
$\Lambda_{\mathrm{eff}}$ is constant. In thermogravity,
however, its evolution is determined by the matter
non-conservation law, without introducing an independent
dynamical field.

For the spatially flat FLRW background considered above,
Eq.~\eqref{eq:effective_Einstein} gives
\begin{equation}
3H^2
=
8\pi G\left(\rho_c+\rho_b+\rho_r\right)
+\Lambda_{\mathrm{eff}}.
\label{eq:effective_Friedmann}
\end{equation}
Differentiating this relation and using
Eqs.~\eqref{eq:running_Hdot} and
\eqref{eq:running_rhoc}-\eqref{eq:running_rhor},
we find
\begin{equation}
\dot{\Lambda}_{\mathrm{eff}}
=
-8\pi G\Gamma(H)\rho_c.
\label{eq:Lambda_eff_dot}
\end{equation}
The geometrical contribution therefore decreases
for positive $\Gamma(H)$ and increases for negative
$\Gamma(H)$, assuming a positive cold dark matter density.
A change in the sign of the non-conservation rate can
consequently reverse its evolution.

It is convenient to associate this contribution with
an effective dark-energy density and pressure,
defined by
\begin{equation}
\rho_{\mathrm{DE}}
\equiv
\frac{\Lambda_{\mathrm{eff}}}{8\pi G},
\qquad
p_{\mathrm{DE}}
\equiv
-\rho_{\mathrm{DE}}.
\label{eq:geometric_density_pressure}
\end{equation}
The reconstructed gravitational equations then take
the familiar Friedmann form
\begin{equation}
H^2
=
\frac{8\pi G}{3}
\left(
\rho_c+\rho_b+\rho_r+\rho_{\mathrm{DE}}
\right),
\label{eq:effective_DE_Friedmann}
\end{equation}
together with the acceleration equation
\begin{equation}
\frac{\ddot a}{a}
=
-\frac{4\pi G}{3}
\left(
\rho_b+\rho_c+2\rho_r-2\rho_{\mathrm{DE}}
\right).
\label{eq:effective_acceleration}
\end{equation}
Thus, a sufficiently positive geometrical contribution
can support accelerated expansion. Its representation
as an effective fluid, however, does not imply the
existence of a fundamental dark-energy component.

From Eqs.
\eqref{eq:Lambda_eff_dot} and \eqref{eq:geometric_density_pressure}, we obtain
\begin{equation}
\dot{\rho}_{\mathrm{DE}}
=
-\Gamma(H)\rho_c.
\label{eq:geometric_conservation}
\end{equation}
Introducing
\begin{equation}
Q(H,\rho_c)\equiv\Gamma(H)\rho_c,
\label{eq:effective_interaction_definition}
\end{equation}
the evolution of cold dark matter and effective
dark energy can be written as
\begin{align}
\dot{\rho}_c+3H\rho_c
&=Q,
\label{eq:effective_interaction_cdm}\\
\dot{\rho}_{\mathrm{DE}}
+3H\left(\rho_{\mathrm{DE}}+p_{\mathrm{DE}}\right)
&=-Q.
\label{eq:effective_interaction_DE}
\end{align}
These equations coincide formally with those of
an interacting dark energy, with $Q>0$
corresponding to energy transfer from the effective
dark energy to cold dark matter, and $Q<0$
to the opposite direction  \cite{Amendola:1999er,Wang:2016lxa,Wang:2024vmw,
Amendola:2000uh,Billyard:2000bh,
Zimdahl:2001ar,Wetterich:1994bg,Cai:2004dk,
delCampo:2008sr, Kumar:2019wfs,Pan:2019gop,
Khoury:2025txd}.

The interpretation of this correspondence requires
some care. In conventional interacting dark-energy
models, the interaction is generally introduced
as a coupling between two physical components.
Here, the non-conservation of cold dark matter
originates from thermogravity, while the effective
dark-energy contribution follows from the
reconstruction of the trace-free gravitational
equations. The interaction term is therefore
determined by the non-conservation law, rather
than introduced arbitrarily, and this may be considered as an advantage of the
present construction.

The reconstructed description also has a conserved
total effective energy-momentum tensor. Indeed, defining
$T_{\mu\nu}^{\mathrm{DE}}
\equiv-\rho_{\mathrm{DE}}g_{\mu\nu}$,
Eq.~\eqref{eq:Lambda_eff_divergence} implies
\begin{equation}
\nabla^\mu
\left(
T_{\mu\nu}+T_{\mu\nu}^{\mathrm{DE}}
\right)=0.
\label{eq:effective_total_conservation}
\end{equation}
This identity does not restore conservation of the
original matter energy-momentum tensor or eliminate
the preferred thermodynamic frame. Furthermore, we mention that
although the homogeneous background admits an
interacting-vacuum representation, this correspondence
does not by itself establish equivalence with a
conventional interacting dark-energy theory at
the level of cosmological perturbations.

For the subsequent analysis, we introduce the
dimensionless geometrical density
\begin{equation}
\lambda(N)
\equiv
\frac{\Lambda_{\mathrm{eff}}(N)}{3H_0^2}
=
\frac{8\pi G\rho_{\mathrm{DE}}(N)}{3H_0^2}.
\label{eq:lambda_definition}
\end{equation}
Using $\xi(E)=\Gamma(H)/H$, we obtain
$
\frac{d\lambda}{dN}
=
-\Omega_{c0}\xi(E)f_c(N)$,
which can be formally integrated as
\begin{equation}
\lambda(N)
=
\lambda_0
-
\Omega_{c0}
\int_0^N
\xi\big(E(\tilde N)\big)
f_c(\tilde N)\,d\tilde N,
\label{eq:lambda_integral}
\end{equation}
where $\lambda_0=\lambda(0)$.
Hence, the reconstructed Friedmann equation becomes
\begin{equation}
E^2(N)
=
\Omega_{b0}e^{-3N}
+\Omega_{r0}e^{-4N}
+\Omega_{c0}f_c(N)
+\lambda(N).
\label{eq:effective_Friedmann_dimensionless}
\end{equation}
Imposing the present-time normalization
$E(0)=f_c(0)=1$, we find
$
\lambda_0
=
1-\Omega_{m0}-\Omega_{r0}$,
where $\Omega_{m0}=\Omega_{b0}+\Omega_{c0}$.
The normalization of the geometrical contribution
is therefore fixed once the present matter and
radiation density parameters are specified.
It should be understood as an integration condition
associated with the missing trace, rather than
an independent prediction of the non-conservation law.
Thus, although constant vacuum-energy contributions
decouple from the trace-free gravitational equations,
the observed magnitude of the effective geometrical
contribution is not independently predicted by the
theory. In this sense, thermogravity reframes the
cosmological-constant problem rather than providing
a complete solution to it.

The cosmological density parameters can then be expressed as
\begin{align}
\Omega_b(N)
&=
\frac{\Omega_{b0}e^{-3N}}{E^2(N)},
\label{eq:Omega_b_evolution}\\
\Omega_c(N)
&=
\frac{\Omega_{c0}f_c(N)}{E^2(N)},
\label{eq:Omega_c_evolution}\\
\Omega_r(N)
&=
\frac{\Omega_{r0}e^{-4N}}{E^2(N)},
\label{eq:Omega_r_evolution}\\
\Omega_{\mathrm{DE}}(N)
&=
\frac{\lambda(N)}{E^2(N)},
\label{eq:Omega_DE_evolution}
\end{align}
and satisfy
\begin{equation}
\Omega_b+\Omega_c+\Omega_r+\Omega_{\mathrm{DE}}=1.
\label{eq:effective_density_constraint}
\end{equation}

Although the  initial geometrical contribution
has a vacuum-like pressure,  the resulting effective energy density is generally
time dependent. In particular, the  intrinsic equation-of-state
parameter is
\begin{equation}
w_{\mathrm{DE}}
\equiv
\frac{p_{\mathrm{DE}}}{\rho_{\mathrm{DE}}}
=-1,
\label{eq:DE_intrinsic_eos}
\end{equation}
wherever $\rho_{\mathrm{DE}}\neq0$.
Nevertheless, to characterize the evolution of its energy density, instead of
(\ref{eq:effective_interaction_DE})
we may   introduce an effective equation-of-state
parameter through
\begin{equation}
\dot{\rho}_{\mathrm{DE}}
+
3H\left(1+w_{\mathrm{DE}}^{\mathrm{eff}}\right)
\rho_{\mathrm{DE}}
=0.
\label{eq:DE_effective_conservation}
\end{equation}
Comparing this expression with
Eq.~\eqref{eq:geometric_conservation} yields
\begin{equation}
w_{\mathrm{DE}}^{\mathrm{eff}}
=
-1+
\frac{\Gamma(H)\rho_c}
{3H\rho_{\mathrm{DE}}},
\label{eq:DE_effective_eos}
\end{equation}
or, equivalently,
\begin{equation}
w_{\mathrm{DE}}^{\mathrm{eff}}(N)
=
-1+
\frac{\Omega_{c0}\xi(E)f_c(N)}
{3\lambda(N)}.
\label{eq:DE_effective_eos_dimensionless}
\end{equation}

For $\rho_{\mathrm{DE}}>0$, a positive
non-conservation rate leads to
$w_{\mathrm{DE}}^{\mathrm{eff}}>-1$,
corresponding to quintessence-like behavior,
whereas a negative rate gives
$w_{\mathrm{DE}}^{\mathrm{eff}}<-1$,
corresponding to an effective phantom regime.
A sign change in $\Gamma(H)$ can therefore
produce a smooth crossing of the phantom divide,
provided that the geometrical density remains
nonzero. This possibility will be illustrated
for the two running prescriptions considered
in the following section.

We emphasize that the effective phantom regime
does not require a phantom field or imply
super-acceleration. Indeed, Eq.~\eqref{eq:running_Hdot}
gives $\dot H\leq0$ for non-negative matter
and radiation densities. Thus,
$w_{\mathrm{DE}}^{\mathrm{eff}}<-1$
characterizes the evolution of the reconstructed
geometrical density, rather than a violation
of the null energy condition by the physical
matter sector. Moreover, if $\rho_{\mathrm{DE}}$
vanishes at a finite redshift,
$w_{\mathrm{DE}}^{\mathrm{eff}}$ may diverge
even when the background expansion remains
regular. Such a divergence reflects the
effective-fluid parametrization and does not,
by itself, represent a physical singularity.

Finally, the cosmological role of the geometrical
contribution depends on both the non-conservation
rate and its integration conditions.
For example, if $\Gamma(H)>0$ and
$\rho_{\mathrm{DE}}$ initially vanishes,
Eq.~\eqref{eq:geometric_conservation} implies
that it becomes non-positive at later times,
and hence cannot support accelerated expansion.
In the conservation limit $\Gamma(H)=0$,
the geometrical density is constant and the
standard $\Lambda$CDM background is recovered
for the corresponding integration conditions.
For nonzero $\Gamma(H)$, its evolution is
directly linked to that of cold dark matter.
These relations provide the basis for the
cosmological analysis of the two models
introduced in the following section.

\section{Cosmological dynamics}
\label{sec:cosmological_dynamics}

In this section, we investigate the cosmological implications
of thermogravity with a running non-conservation scale.
We first introduce two specific prescriptions for the
non-conservation rate and examine their analytical and
numerical background evolution, focusing on the matter
densities and the effective geometric dark-energy sector.
We then perform a dynamical-system analysis to identify
the relevant critical points, their stability, and the
possible late-time cosmological attractors.

\subsection{Specific models}
\label{subsec:specific_models}

Having established the cosmological equations for a general
non-conservation rate $\Gamma(H)$, we now investigate two
specific prescriptions for its dependence on the Hubble
parameter. The first assumes a rate proportional to $H$,
while the second introduces an additional parameter
controlling its evolution through an inverse-Hubble
dependence. These choices allow us to examine different
non-conservation histories within the same thermogravity
framework.

\textit{Model I: Hubble-proportional non-conservation.}
We first consider
\begin{equation}
\Gamma(H)=\xi_0 H,
\label{eq:Gamma_model_I}
\end{equation}
where $\xi_0$ is a dimensionless constant.
The corresponding non-conservation scale is
$L_2(H)=1/(\xi_0 H)$.
The cold dark matter continuity equation therefore becomes
\begin{equation}
\dot{\rho}_c+3H\rho_c=\xi_0 H\rho_c.
\label{eq:rhoc_model_I}
\end{equation}
Positive values of $\xi_0$ correspond to matter production,
whereas negative values describe matter depletion.
For $\xi_0=0$, standard matter conservation is recovered,
together with the $\Lambda$CDM background for the
corresponding integration conditions.

An important feature of this prescription is that the
relative non-conservation rate $
\frac{\Gamma(H)}{H}=\xi_0$   remains constant
throughout the cosmological evolution. Consequently,
although the absolute rate changes with $H$, its
contribution to the cold dark matter dilution law is
not suppressed at early times. This distinguishes Model I
from prescriptions in which $\Gamma/H$ decreases as the
Hubble parameter increases. Moreover, the resulting
continuity equation has the same mathematical structure
as that of interacting-vacuum models with
$Q\propto H\rho_c$, although its origin here is the
thermogravity non-conservation law. As we will show below,
in the pressureless limit this model is also related
to the quadratic running-vacuum description.

\textit{Model II: Linear running non-conservation.}
As a second possibility, we consider
\begin{equation}
\Gamma(H)
=
H_0\left[
\xi_0+\xi_1
\left(1-\frac{H_0}{H}\right)
\right],
\label{eq:Gamma_model_II}
\end{equation}
where $\xi_0$ and $\xi_1$ are dimensionless parameters.
The parameter $\xi_0$ determines the present value
$\Gamma(H_0)=\xi_0 H_0$, while $\xi_1$ controls
the running of the non-conservation rate.
This prescription can be regarded as a first-order
expansion of a general function $\Gamma(H)$ in the
variable $1-H_0/H$ around the present epoch.
It provides a simple phenomenological extension of
the original constant-scale thermogravity model,
without specifying a particular power-law behavior
over the entire cosmological history.

The corresponding dimensionless non-conservation rate is
\begin{equation}
\frac{\Gamma(H)}{H}
=
\frac{\xi_0+\xi_1}{E}
-\frac{\xi_1}{E^2},
\label{eq:Gamma_over_H_model_II}
\end{equation}
where $E=H/H_0$. Accordingly, the cold dark matter
continuity equation reads
\begin{equation}
\dot{\rho}_c+3H\rho_c
=
H_0\left[
\xi_0+\xi_1
\left(1-\frac{H_0}{H}\right)
\right]\rho_c.
\label{eq:rhoc_model_II}
\end{equation}
For $\xi_1=0$, we recover the original constant-rate
thermogravity model of \cite{Lee:2026ayb}, whereas $\xi_0=\xi_1=0$ gives
standard matter conservation and the $\Lambda$CDM
background.

Unlike Model I, the relative non-conservation rate
vanishes in the high-curvature limit. Indeed, for
$H/H_0\rightarrow\infty$, we obtain
\begin{equation}
\Gamma(H)\longrightarrow H_0(\xi_0+\xi_1),
\qquad
\frac{\Gamma(H)}{H}\longrightarrow0.
\label{eq:Gamma_model_II_early}
\end{equation}
Thus, non-conservation becomes negligible relative to the
cosmological expansion at sufficiently early times, and
the local cold dark matter dilution law approaches its
standard form. This behavior may facilitate agreement
with early-Universe observations, although it does not
by itself guarantee consistency with Big Bang
Nucleosynthesis or cosmic microwave background
constraints. In particular, the asymptotic normalization
may differ from that of conserved matter, owing to the
cumulative effects of non-conservation during the
subsequent evolution.

Model II also allows the non-conservation rate to
change sign. Setting $\Gamma(H_\times)=0$, we find
\begin{equation}
\frac{H_\times}{H_0}
=
\frac{\xi_1}{\xi_0+\xi_1},
\label{eq:Gamma_model_II_crossing}
\end{equation}
provided $\xi_0+\xi_1\neq0$.
Since $H$ decreases with cosmic time for non-negative
matter and radiation densities, a sign change in
the past, corresponding to $H_\times>H_0$, occurs when
\begin{equation}
\xi_0(\xi_0+\xi_1)<0.
\label{eq:Gamma_model_II_crossing_condition}
\end{equation}
Such a transition reverses the direction of the
effective energy exchange between cold dark matter
and the reconstructed geometrical contribution, resembling the interacting
dark-energy scenarios with sign-changing interaction forms
\cite{Cai:2009ht,Wei:2010fz,Pan:2019jqh}.
It can also induce a crossing of
$w_{\mathrm{DE}}^{\mathrm{eff}}=-1$, provided that
the geometrical dark-energy density remains nonzero.

The two prescriptions therefore exhibit different
non-conservation histories. Model I maintains a
constant ratio $\Gamma/H$, modifying the cold dark
matter dilution law throughout the expansion, whereas
Model II approaches standard dilution at early times
and allows the direction of energy exchange to change.
Neither prescription is uniquely determined by the
microscopic thermodynamic construction, and their
cosmological viability must ultimately be assessed
through the resulting background evolution and
observational constraints. We now examine their
analytical and numerical properties.

\subsection{Analytical and numerical background evolution}
\label{subsec:background_evolution}

We now examine the background evolution of the two
non-conservation prescriptions introduced above.
Model I admits an exact solution, including baryons and
radiation, whereas Model II requires numerical integration
of the cosmological equations.

Independently of the particular form of $\Gamma(H)$,
the deceleration parameter can be expressed as
\begin{eqnarray}
&&
\!\!\!\!\!\!\!\!\!\!\!\!\!\!\!\!\!
q(z)\equiv-1-\frac{\dot H}{H^2}\nonumber\\
&&\!\!\!\!\!\!\!\!\!\!\!\!\!\!\!\!\!
=-1+\frac{3\Omega_{b0}(1+z)^3
+3\Omega_{c0}f_c(z)+4\Omega_{r0}(1+z)^4}
{2E^2(z)}.
\label{eq:background_q_general}
\end{eqnarray}
At the present epoch, this reduces to
$
q_0=-1+\frac{3}{2}\Omega_{m0}+2\Omega_{r0}$,
which is independent of the non-conservation parameters
for fixed present-day matter and radiation densities.
Thus, the two models modify the redshift dependence of
the deceleration parameter, while predicting the same
$q_0$ when their present density parameters coincide.

A useful analytical relation can also be obtained in
the dust-only limit. In a spatially flat background
containing cold dark matter, the trace-free gravitational
equation gives $\dot H=-4\pi G\rho_c$. Differentiating
this expression and using the modified continuity
equation, for non-vacuum solutions with $\dot H\neq0$ we find
\begin{equation}
\frac{d\dot H}{dH}=\Gamma(H)-3H.
\label{eq:dust_firstorder}
\end{equation}
This relation provides a first integral for an arbitrary
non-conservation rate and will be useful for examining
the analytical properties of the two models.

\subsubsection{Model I: Hubble-proportional non-conservation}
\label{subsubsec:background_model_I}

For $\Gamma(H)=\xi_0H$, the cold dark matter
continuity equation can be integrated directly.
Using $N=\ln a$, we obtain
\begin{equation}
\frac{d\rho_c}{dN}=-(3-\xi_0)\rho_c,
\qquad
\rho_c(z)=\rho_{c0}(1+z)^{3-\xi_0}.
\label{eq:model_I_rhoc_solution}
\end{equation}
Hence, $f_c(z)=(1+z)^{3-\xi_0}$, showing that
positive values of $\xi_0$ reduce the past cold dark
matter density relative to the standard dust scaling
for a fixed present-day normalization. Negative
values produce the opposite effect.

Substituting this solution into
Eq.~\eqref{eq:running_E2_redshift} and imposing
$E(0)=1$, we obtain
\begin{align}
E^2(z)=&\,1+\Omega_{b0}\left[(1+z)^3-1\right]
+\Omega_{r0}\left[(1+z)^4-1\right]\nonumber\\
&+\frac{3\Omega_{c0}}{3-\xi_0}
\left[(1+z)^{3-\xi_0}-1\right],
\label{eq:model_I_E2_exact}
\end{align}
valid for $\xi_0\neq3$. For $\xi_0=3$, the last
term is replaced by $3\Omega_{c0}\ln(1+z)$.
Thus, the expansion history $H(z)=H_0E(z)$ is
determined analytically and reduces to the
spatially flat $\Lambda$CDM result when $\xi_0=0$.

The geometrical dark-energy contribution can also
be reconstructed explicitly. From the effective
Friedmann constraint, we find
\begin{equation}
\lambda(z)=\lambda_0+
\frac{\xi_0\Omega_{c0}}{3-\xi_0}
\left[(1+z)^{3-\xi_0}-1\right],
\qquad (\xi_0\neq3),
\label{eq:model_I_lambda_exact}
\end{equation}
where $\lambda_0=1-\Omega_{m0}-\Omega_{r0}$.
The corresponding density parameters are
\begin{equation}
\Omega_c(z)=\frac{\Omega_{c0}(1+z)^{3-\xi_0}}{E^2(z)},
\qquad
\Omega_{\mathrm{DE}}(z)=\frac{\lambda(z)}{E^2(z)},
\label{eq:model_I_omegas}
\end{equation}
with the baryon and radiation contributions following
from their standard conservation laws.
Finally, the effective geometric dark-energy
equation of state becomes
\begin{equation}
w_{\mathrm{DE}}^{\mathrm{eff}}(z)
=-1+\frac{\xi_0\Omega_{c0}(1+z)^{3-\xi_0}}
{3\lambda(z)}.
\label{eq:model_I_w_eff_exact}
\end{equation}
These expressions determine the complete homogeneous
background evolution of Model I.

\begin{figure}[ht!]
\centering
\includegraphics[width=1.0\linewidth]{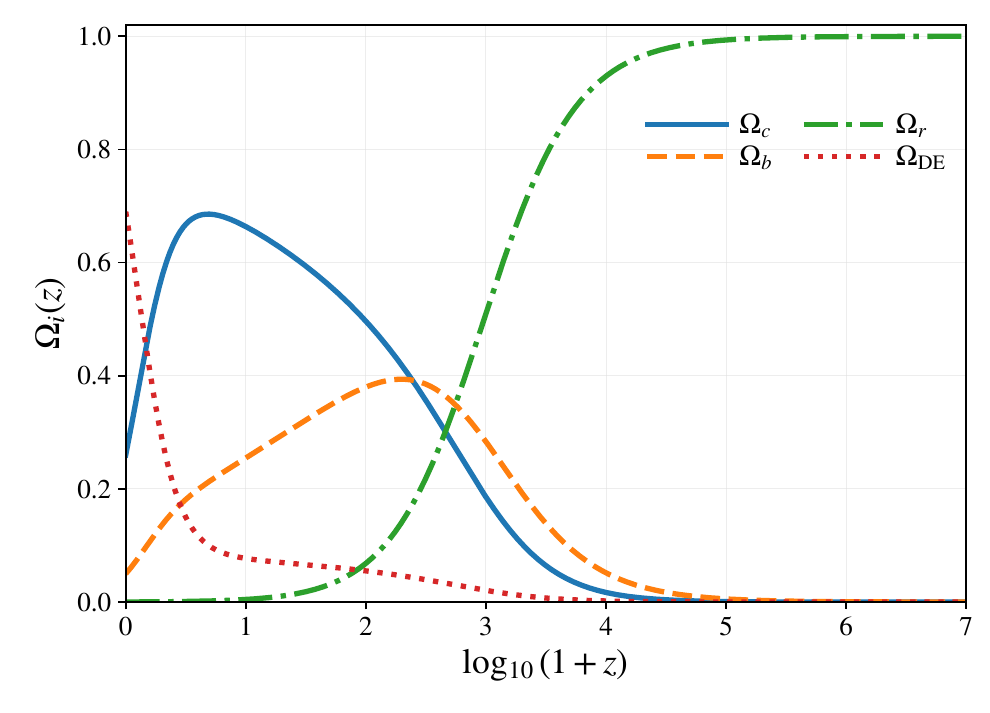}
\caption{\textit{Evolution of the density parameters
$\Omega_c$, $\Omega_b$, $\Omega_r$, and
$\Omega_{\mathrm{DE}}$ in Model I of (\ref{eq:Gamma_model_I}), for the illustrative
choice $\xi_0=0.30$. We adopt $\Omega_{m0}=0.31$,
$\Omega_{b0}=0.05$, and $\Omega_{r0}=9\times10^{-5}$.
The geometric dark-energy contribution is reconstructed
from the trace-free gravitational equations.}}
\label{fig:model_I_densities}
\end{figure}

In Fig.~\ref{fig:model_I_densities}, we present the
evolution of the density parameters for a representative
positive non-conservation rate. The Universe evolves
from an early radiation-dominated regime through
matter domination towards a late epoch dominated
by the effective geometrical contribution.
The modified dilution of cold dark matter affects
its relative abundance, while baryons and radiation
retain their standard conservation laws.
At sufficiently high redshifts, radiation dominates
provided that the cold dark matter contribution
grows more slowly toward the past, as occurs for
$\xi_0>-1$. For more negative values, the standard
early radiation regime is not generically recovered.

\begin{figure}[ht!]
\centering
\includegraphics[width=1.0\linewidth]{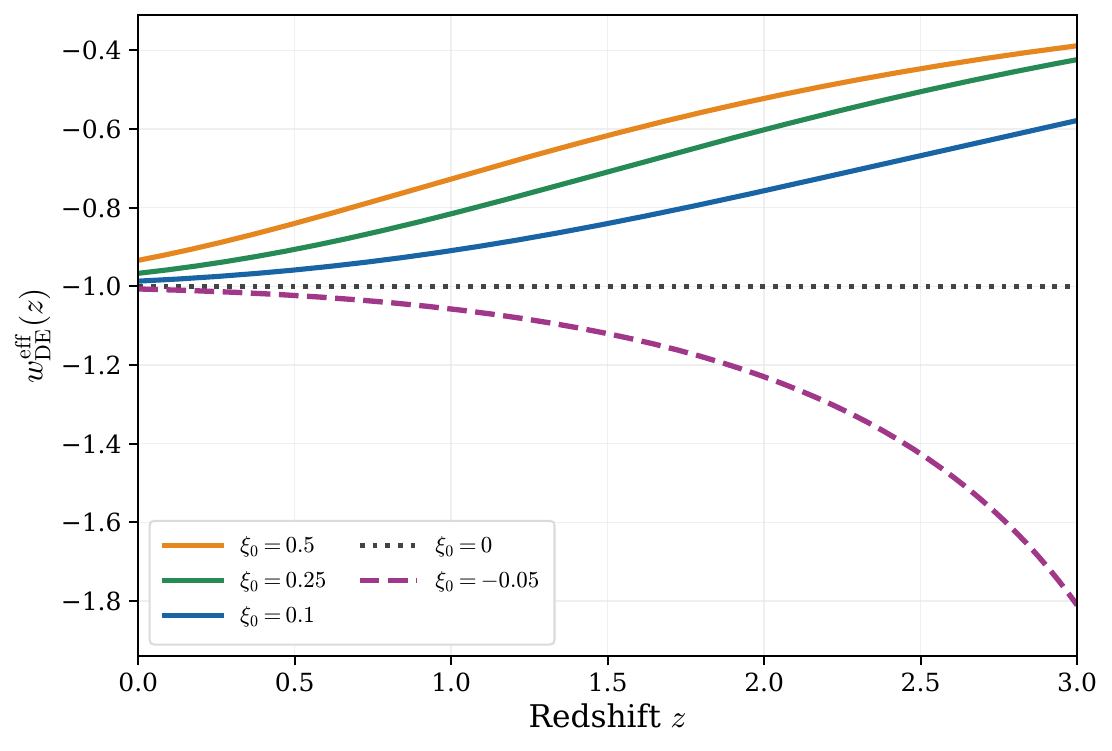}
\caption{\textit{Evolution of the effective geometric
dark-energy equation-of-state parameter
$w_{\mathrm{DE}}^{\mathrm{eff}}(z)$ in Model I of (\ref{eq:Gamma_model_I}),
for five illustrative values of $\xi_0$.
Positive values correspond to quintessence-like
behavior, while negative values lead to an
effective phantom regime. The choice $\xi_0=0$
recovers the $\Lambda$CDM scenario. We adopt
$\Omega_{m0}=0.31$, $\Omega_{b0}=0.05$,
and $\Omega_{r0}=9\times10^{-5}$.}}
\label{fig:model_I_w_eff}
\end{figure}

Figure~\ref{fig:model_I_w_eff} displays the evolution
of $w_{\mathrm{DE}}^{\mathrm{eff}}$ for five values
of $\xi_0$. Positive values give rise to
quintessence-like behavior, with
$w_{\mathrm{DE}}^{\mathrm{eff}}>-1$, whereas
negative values lead to an effective phantom regime,
$w_{\mathrm{DE}}^{\mathrm{eff}}<-1$.
The departure from $-1$ becomes more pronounced
as the magnitude of the non-conservation parameter
increases. For $\xi_0=0$, the effective equation
of state remains equal to $-1$, reproducing
the $\Lambda$CDM limit.

We note that the geometrical density can also vanish
at a finite redshift for suitable parameter values.
For $\xi_0\neq0,3$, the condition
$\lambda(z_\times)=0$ gives
\begin{equation}
(1+z_\times)^{3-\xi_0}
=1-\frac{(3-\xi_0)\lambda_0}
{\xi_0\Omega_{c0}},
\label{eq:model_I_lambda_crossing}
\end{equation}
provided that a real solution exists in the relevant
redshift interval. At such a crossing,
$w_{\mathrm{DE}}^{\mathrm{eff}}$ may diverge even
though the expansion history remains regular.

The dust-only limit provides a further analytical
connection with running-vacuum cosmologies.
Integrating Eq.~\eqref{eq:dust_firstorder} for
$\Gamma(H)=\xi_0H$, we obtain
\begin{equation}
\dot H=-\frac{3-\xi_0}{2}H^2+C_{\mathrm I},
\label{eq:model_I_dust_first_integral}
\end{equation}
where $C_{\mathrm I}$ is determined by the
integration conditions. Since
$\Lambda_{\mathrm{eff}}=3H^2+2\dot H$ in this limit,
it follows that
\begin{equation}
\rho_{\mathrm{DE}}(H)=\rho_{\mathrm{DE},0}
+\frac{\xi_0}{8\pi G}(H^2-H_0^2).
\label{eq:model_I_running_vacuum}
\end{equation}
Thus, Model I reproduces the quadratic
running-vacuum form in the dust-only background,
with the identification $\nu=\xi_0/3$.
This relation does not extend unchanged to the
complete system, since baryons and radiation
also contribute to the evolution of $H$.

\subsubsection{Model II: Linear running non-conservation}
\label{subsubsec:background_model_II}

For Model II, the relative non-conservation rate
is given by Eq.~\eqref{eq:Gamma_over_H_model_II}.
Unlike Model I, the cold dark matter evolution
depends explicitly on the expansion history,
and the full background must therefore be
obtained numerically.

Substituting the Model II prescription into
Eqs.~\eqref{eq:running_E2_redshift} and
\eqref{eq:running_fc_redshift}, we obtain
\begin{align}
\frac{dE^2}{dz}
&=3\Omega_{b0}(1+z)^2
+4\Omega_{r0}(1+z)^3
+\frac{3\Omega_{c0}f_c(z)}{1+z},
\label{eq:model_II_E2_redshift}\\
\frac{df_c}{dz}
&=\frac{f_c(z)}{1+z}
\left[3-\frac{\xi_0+\xi_1}{E}
+\frac{\xi_1}{E^2}\right],
\label{eq:model_II_fc_redshift}
\end{align}
with $E(0)=f_c(0)=1$.
The dimensionless geometrical density is then
reconstructed from
\begin{equation}
\lambda(z)=E^2(z)-\Omega_{b0}(1+z)^3
-\Omega_{r0}(1+z)^4-\Omega_{c0}f_c(z),
\label{eq:model_II_lambda_reconstruction}
\end{equation}
or, equivalently, by integrating
\begin{equation}
\frac{d\lambda}{dz}
=\frac{\Omega_{c0}f_c(z)}{1+z}
\left[\frac{\xi_0+\xi_1}{E}
-\frac{\xi_1}{E^2}\right],
\qquad
\lambda(0)=\lambda_0.
\label{eq:model_II_lambda_redshift}
\end{equation}
The agreement between these expressions provides
a useful consistency check of the numerical solution.
The effective geometrical dark-energy equation of
state follows as
\begin{equation}
w_{\mathrm{DE}}^{\mathrm{eff}}(z)
=-1+\frac{\Omega_{c0}f_c(z)}{3\lambda(z)}
\left[\frac{\xi_0+\xi_1}{E(z)}
-\frac{\xi_1}{E^2(z)}\right].
\label{eq:model_II_w_eff_background}
\end{equation}

The early-time behavior differs from that of
Model I. For $E\gg1$, the relative
non-conservation rate approaches zero, and
the local cold dark matter dilution law tends
to its standard form. More precisely, provided
that the integrated non-conservation history
converges toward the past, one obtains
$f_c(z)\sim A_c(1+z)^3$, where $A_c$ is
a constant determined by the evolution.
Thus, the standard scaling is recovered
asymptotically, although its normalization
may differ from that of conserved cold dark
matter with the same present-day density.

An additional analytical result follows from
the dust-only first integral. Substituting
Eq.~\eqref{eq:Gamma_model_II} into
Eq.~\eqref{eq:dust_firstorder}, we obtain
\begin{equation}
\dot H=-\frac{3}{2}H^2
+H_0(\xi_0+\xi_1)H
-\xi_1H_0^2\ln\!\left(\frac{H}{H_0}\right)
+C_{\mathrm{II}},
\label{eq:model_II_dust_first_integral}
\end{equation}
where $C_{\mathrm{II}}$ is fixed by the
integration conditions. This expression
provides an implicit analytical description
of the dust-only evolution, while the complete
background with baryons and radiation is
determined from the numerical system above.

The density parameters in Model II exhibit
a qualitatively similar cosmological sequence
to that shown for Model I in
Fig.~\ref{fig:model_I_densities}, with their
relative contributions modified by the running
non-conservation rate. A more distinctive
feature of Model II is the possibility that
$\Gamma(H)$ changes sign, reversing the
direction of energy exchange between cold dark
matter and the effective geometrical sector.
Provided that $\rho_{\mathrm{DE}}$ remains
nonzero, such a transition produces a crossing
of the effective phantom divide.

\begin{figure}[ht!]
\centering
\includegraphics[width=1.0\linewidth,height=0.68\linewidth]{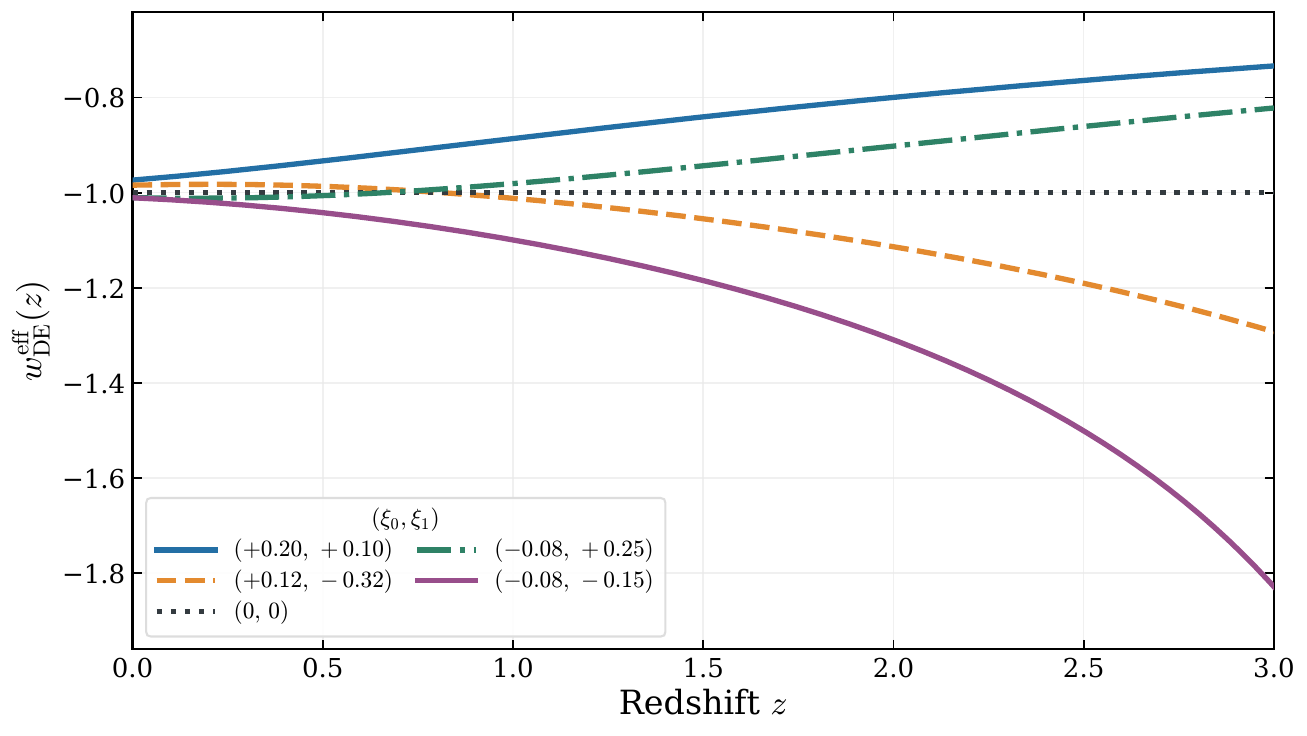}
\caption{\textit{Evolution of the effective geometric
dark-energy equation-of-state parameter
$w_{\mathrm{DE}}^{\mathrm{eff}}(z)$ in Model II of (\ref{eq:Gamma_model_II}),
for five illustrative combinations of the
non-conservation parameters $(\xi_0,\xi_1)$.
Depending on these parameters, the model exhibits
quintessence-like or phantom-like behavior,
as well as smooth crossings of the phantom
divide in either direction. The choice
$(\xi_0,\xi_1)=(0,0)$ reproduces $\Lambda$CDM.
The geometrical dark-energy density remains
positive throughout the plotted interval.
We adopt $\Omega_{m0}=0.31$, $\Omega_{b0}=0.05$,
and $\Omega_{r0}=9\times10^{-5}$.}}
\label{fig:model_II_w_eff}
\end{figure}

This behavior is illustrated in
Fig.~\ref{fig:model_II_w_eff}, where we present
$w_{\mathrm{DE}}^{\mathrm{eff}}(z)$ for five
combinations of $(\xi_0,\xi_1)$. Depending on
the parameters, the effective equation of state
can remain above or below $-1$, or cross this
value during the cosmological evolution.
In particular, the choices
$(\xi_0,\xi_1)=(0.12,-0.32)$ and
$(-0.08,0.25)$ produce smooth crossings
in opposite directions, at approximately
$z\simeq0.82$ and $z\simeq0.66$, respectively.
The geometrical dark-energy density remains
positive over the plotted interval for all
the displayed parameter combinations.
The $\Lambda$CDM limit is recovered for
$(\xi_0,\xi_1)=(0,0)$.

An interesting property of the two models is
that their present-day effective dark-energy
equation of state is determined by the same
expression. Since $\Gamma(H_0)/H_0=\xi_0$
in both cases, we obtain
\begin{equation}
w_{\mathrm{DE},0}^{\mathrm{eff}}
=-1+\frac{\xi_0\Omega_{c0}}
{3(1-\Omega_{m0}-\Omega_{r0})}.
\label{eq:effective_DE_eos_today}
\end{equation}
Thus, for fixed $\xi_0$ and cosmological
density parameters, the additional running
parameter $\xi_1$ in Model II affects the
redshift evolution of the effective equation
of state without changing its present-day value.

We mention here that both models can approach a late-time
de Sitter regime, characterized by
$H\rightarrow H_\star>0$ and vanishing
matter and radiation densities.
The existence of such a regime, however,
does not guarantee that it is reached by every
cosmological solution. Its stability depends
on the behavior of the non-conservation rate
near the asymptotic state, while the value
of $H_\star$ is selected by the integration
conditions and the cosmological trajectory.
These properties will be examined systematically
through the dynamical-system analysis of the
following subsection.

Finally, concerning the early-Universe evolution, both
models can approach the standard cosmological behavior
for suitable choices of their non-conservation parameters.
Nevertheless, a complete assessment of their consistency
with early-Universe observations, particularly regarding
matter-radiation equality, recombination, and the sound
horizon, requires complementary constraints from Big Bang
Nucleosynthesis and the cosmic microwave background.
Such an analysis lies beyond the scope of the present work.

\subsection{Dynamical-system formulation and critical points}
\label{subsec:dynamical_system}

To complement the background analysis, we investigate the
critical points and their stability \cite{Copeland:1997et,Bahamonde:2017ize},
focusing on the
cosmological regimes relevant to the two models.
The phase-space structure of basic constant-scale thermogravity model
has been examined in \cite{Brereton:2026mki}.
Here, we employ a unified autonomous formulation that
accommodates both running prescriptions and allows
their qualitative properties to be compared directly.

We introduce the dimensionless variables
\begin{equation}
x\equiv\Omega_c,\qquad y\equiv\Omega_b,\qquad
r\equiv\Omega_r,\qquad u\equiv\frac{H_0}{H},
\label{eq:ds_variables}
\end{equation}
where $x,y,r\geq0$ and $u>0$ for expanding solutions.
The effective geometrical density parameter is
$\Omega_{\mathrm{DE}}=1-x-y-r$. Since this contribution
is not necessarily positive, the physical phase space
is not generally restricted to $x+y+r\leq1$.
Using $N=\ln a$ as the evolution variable, with a
prime denoting $d/dN$, we further define
\begin{equation}
\mathcal{A}\equiv3(x+y)+4r,
\qquad
\mathcal{X}(u)\equiv\frac{\Gamma(H)}{H}.
\label{eq:ds_auxiliary}
\end{equation}
For the two models, we have
\begin{equation}
\mathcal{X}(u)=
\begin{cases}
\xi_0, & \text{Model I},\\
(\xi_0+\xi_1)u-\xi_1u^2, & \text{Model II}.
\end{cases}
\label{eq:ds_model_functions}
\end{equation}
The cosmological equations then take the autonomous form
\begin{align}
x'&=x[-3+\mathcal{X}(u)+\mathcal{A}],
\label{eq:ds_xprime}\\
y'&=y(-3+\mathcal{A}),
\label{eq:ds_yprime}\\
r'&=r(-4+\mathcal{A}),
\label{eq:ds_rprime}\\
u'&=\frac{u}{2}\mathcal{A}.
\label{eq:ds_uprime}
\end{align}
The deceleration parameter and the effective total
equation-of-state parameter are given by
\begin{equation}
q=-1+\frac{\mathcal{A}}{2},
\qquad
w_{\mathrm{tot}}=-1+\frac{\mathcal{A}}{3}.
\label{eq:ds_cosmological_quantities}
\end{equation}
For non-negative matter and radiation densities,
$u'\geq0$, consistently with the monotonic decrease
of $H$ during the expanding evolution. The variable
$u=H_0/H$ also remains regular when the
non-conservation rate changes sign.

The critical configurations follow by setting
the right-hand sides of
Eqs.~\eqref{eq:ds_xprime}-\eqref{eq:ds_uprime}
to zero. Their main properties are summarized
in Table~\ref{tab:ds_points_two_models}.

\begin{table}[ht!]
\centering
\caption{Relevant critical configurations
and their stability properties for Models I and II.
The points $\mathcal{R}$, $\mathcal{B}$, and
$\mathcal{D}$ are common to both models, while
$\mathcal{S}_{\mathrm I}$ and $\mathcal{M}_{\alpha}$
characterize Models I and II, respectively.
The indicated stability properties apply under
the conditions specified below.}
\label{tab:ds_points_two_models}
\begin{tabular}{l c c c}
\hline\hline
Point & $(x,y,r,u)$ & $q$ & Character \\
\hline
$\mathcal{R}$ & $(0,0,1,0)$
& $1$ & Past attractor$^a$ \\
$\mathcal{B}$ & $(0,1,0,0)$
& $1/2$ & Saddle \\
$\mathcal{S}_{\mathrm I}$ &
$(1-\xi_0/3,0,0,0)$
& $(1-\xi_0)/2$ & Saddle$^b$ \\
$\mathcal{M}_\alpha$ &
$(\alpha,1-\alpha,0,0)$
& $1/2$ & Saddle line \\
$\mathcal{D}_{u_\star}$ &
$(0,0,0,u_\star)$
& $-1$ & Attracting$^c$ \\
\hline\hline
\end{tabular}
\par\vspace{2pt}
{\footnotesize
$^a$ For Model I if $\xi_0>-1$; for Model II
independently of its parameters.
$^b$ For $-1<\xi_0<3$, $\xi_0\neq0$.
$^c$ Transversely, if $\mathcal{X}(u_\star)<3$.}
\end{table}

At the high-curvature boundary $u=0$, both models
admit the radiation-dominated critical point
$\mathcal{R}=(0,0,1,0)$, characterized by $q=1$
and total equation-of-state parameter $w_{\mathrm{tot}}=1/3$. The corresponding
eigenvalues are
\begin{equation}
\begin{aligned}
\mathcal{R}:\quad
&\{1+\xi_0,1,4,2\}, &&\text{Model I},\\
&\{1,1,4,2\}, &&\text{Model II}.
\end{aligned}
\label{eq:ds_radiation_eigenvalues}
\end{equation}
Therefore, $\mathcal{R}$ acts as a past attractor
for Model I when $\xi_0>-1$, and for Model II
independently of its parameters. Both models
also admit the pure-baryon configuration
$\mathcal{B}=(0,1,0,0)$, which is a saddle point
with $q=1/2$.

The matter sector exhibits a qualitative difference
between the two prescriptions. For Model II,
$\mathcal{X}(0)=0$, and the system admits the
matter-dominated critical line
\begin{equation}
\mathcal{M}_{\alpha}:
\quad(x,y,r,u)=(\alpha,1-\alpha,0,0),
\qquad 0\leq\alpha\leq1.
\label{eq:ds_matter_line_II}
\end{equation}
Its eigenvalues are $\{3,-1,0,3/2\}$, indicating
a saddle set with a neutral direction along the
critical line. These configurations reproduce
the standard matter-dominated expansion, with
$w_{\mathrm{tot}}=0$ and $q=1/2$.

For Model I, by contrast, a nonzero $\xi_0$
modifies the matter critical structure.
In particular, for $\xi_0<3$, the system admits
the scaling point
\begin{equation}
\mathcal{S}_{\mathrm I}:
\quad(x,y,r,u)=
\left(1-\frac{\xi_0}{3},0,0,0\right),
\label{eq:ds_scaling_I}
\end{equation}
for which $\Omega_{\mathrm{DE}}=\xi_0/3$ and
$
w_{\mathrm{tot}}=-\frac{\xi_0}{3},$
 $
q=\frac{1-\xi_0}{2}$.
The corresponding eigenvalues are
$\{3-\xi_0,-\xi_0,-1-\xi_0,(3-\xi_0)/2\}$.
Thus, for $-1<\xi_0<3$, with $\xi_0\neq0$,
this configuration is a saddle. It represents
a scaling solution involving cold dark matter
and effective geometrical dark energy, located
in the baryon-free boundary of phase space.
Hence, it should not be identified with a
generic mixed baryon and cold dark matter
equilibrium. The standard matter critical
line is recovered when $\xi_0=0$.

At late times, both models admit a continuous
family of de Sitter critical points,
\begin{equation}
\mathcal{D}_{u_\star}:
\quad(x,y,r,u)=(0,0,0,u_\star),
\qquad u_\star>0,
\label{eq:ds_desitter_family}
\end{equation}
where $H_\star=H_0/u_\star$,
$\Omega_{\mathrm{DE}}=1$, and $q=-1$.
The eigenvalues of the linearized system are
\begin{equation}
\left\{
\mathcal{X}(u_\star)-3,-3,-4,0
\right\}.
\label{eq:ds_desitter_eigenvalues}
\end{equation}
The zero eigenvalue corresponds to the
continuous family of equilibria, while the
remaining eigenvalues determine their transverse
stability. Consequently, the de Sitter family
is normally attracting when
$\mathcal{X}(u_\star)<3$, whereas configurations
with $\mathcal{X}(u_\star)>3$ are unstable
against cold dark matter perturbations.
For the two models, the attracting condition becomes
\begin{equation}
\begin{aligned}
&\xi_0<3, &&\text{Model I},\\
&(\xi_0+\xi_1)u_\star-\xi_1u_\star^2<3,
&&\text{Model II}.
\end{aligned}
\label{eq:ds_desitter_stability}
\end{equation}
The limiting case $\mathcal{X}(u_\star)=3$
has an additional zero eigenvalue and requires
a nonlinear treatment, which lies beyond the
scope of the present analysis.

An important feature is that the de Sitter
attractor is not a unique critical point.
Instead, different values of $u_\star$ correspond
to different asymptotic expansion rates.
Each cosmological trajectory approaching the
normally attracting family selects a particular
de Sitter configuration, determined by its
initial conditions and the integration freedom
associated with the missing trace of the
gravitational field equations.

\begin{figure*}[t]
\centering
\includegraphics[width=0.95\textwidth]{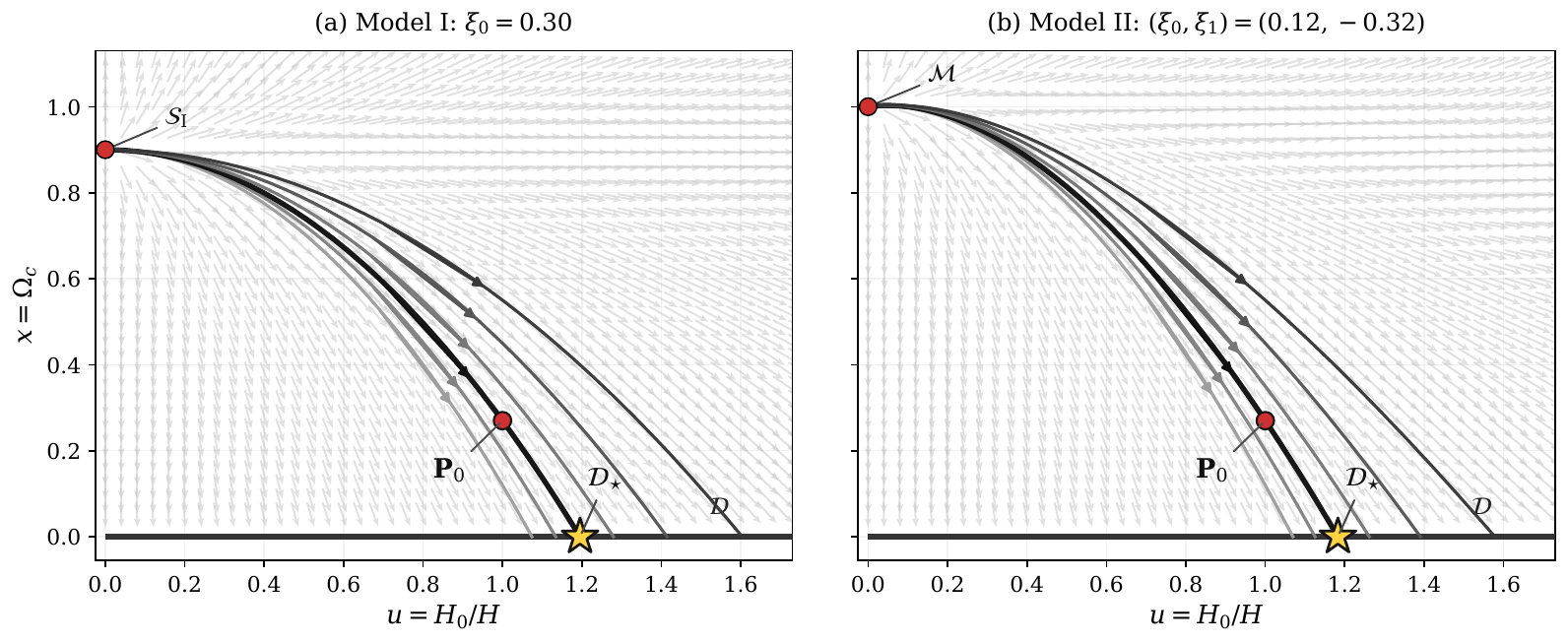}
\caption{\textit{Phase-space trajectories in the
invariant dust-only subspace for (a) Model I of (\ref{eq:Gamma_model_I})
with $\xi_0=0.30$ and (b) Model II of (\ref{eq:Gamma_model_II}) with
$(\xi_0,\xi_1)=(0.12,-0.32)$.
Here $u=H_0/H$ and $x=\Omega_c$.
The red points at $u=0$ indicate the
matter-era configurations
$\mathcal{S}_{\mathrm I}$ and
$\mathcal{M}_{\alpha=1}$, respectively,
while the black horizontal line
represents the de Sitter critical family
$\mathcal{D}$.
The red points $P_0$ correspond to the
illustrative initial conditions
$(u_0,x_0)=(1,0.27)$.
The highlighted black trajectories approach
the particular de Sitter equilibria
$\mathcal{D}_\star$, indicated by yellow stars.
Arrows show the direction of increasing
$N=\ln a$.}}
\label{fig:ds_phase_space}
\end{figure*}

The asymptotic expansion rate can also be related to the
cosmological parameters. In particular, for Model I,
taking the late-time limit of the exact background
solution (\ref{eq:model_I_E2_exact}), we obtain
\begin{equation}
\frac{H_\star^2}{H_0^2}
=1-\Omega_{b0}-\Omega_{r0}
-\frac{3\Omega_{c0}}{3-\xi_0},
\label{eq:Hstar_model_I}
\end{equation}
provided that $\xi_0<3$ and the right-hand side is
positive. For instance, adopting $\Omega_{m0}=0.31$,
$\Omega_{b0}=0.05$, $\Omega_{r0}=9\times10^{-5}$,
and $\xi_0=0.08$, we find $H_\star\simeq0.826H_0$,
which satisfies the stability condition $\xi_0<3$.
For Model II, the corresponding asymptotic rate
can be determined through the numerical integration
of the background equations, subject to the stability
condition (\ref{eq:ds_desitter_stability}). Thus, although the de Sitter critical
points form a continuous family, the cosmological
evolution selects a specific asymptotic expansion rate
for given parameters and integration conditions.

To illustrate the phase-space evolution, we restrict
the system to the invariant dust-only subspace
$y=r=0$. The autonomous equations then reduce to
$x'=x[-3+\mathcal{X}(u)+3x]$ and $u'=3ux/2$.
In Fig.~\ref{fig:ds_phase_space}, we present
representative trajectories for both models.
The trajectories emerge from the vicinity of the
corresponding matter critical configurations
and approach the normally attracting de Sitter
family, represented by the horizontal line $x=0$.
The two phase portraits are qualitatively similar,
since both prescriptions possess the same
de Sitter critical family, although their matter
critical structures and non-conservation histories
differ.

To distinguish the continuous attractor family
from the asymptotic state selected by a particular
solution, we also highlight a trajectory starting
from $P_0$, with $(u_0,x_0)=(1,0.27)$.
For Model I with $\xi_0=0.30$, this trajectory
approaches $\mathcal{D}_\star$ at
$u_\star\simeq1.195$, while for Model II with
$(\xi_0,\xi_1)=(0.12,-0.32)$, the corresponding
value is $u_\star\simeq1.183$.
These values refer to the illustrative dust-only
subspace and are not the asymptotic predictions
of the complete system containing baryons and
radiation. The figure therefore illustrates
the qualitative phase-space structure rather
than a full cosmological evolution through
all successive epochs.

\section{Observational constraints}
\label{sec:observational_constraints}

In this section, we confront the two thermogravity models
with current cosmological observations, in order to constrain
their parameters and assess their observational viability.
We first describe the datasets and statistical methodology,
and then present the resulting parameter constraints and
correlations. Finally, we compare the statistical performance
of the two models with the $\Lambda$CDM paradigm and constant-rate
thermogravity, using standard information criteria.
\subsection{Observational datasets and methodology}
\label{subsec:observational_methodology}

Having investigated the background evolution and dynamical
properties of the two thermocosmological models, we now confront
them with cosmological observations. We employ a combination
of complementary late-time probes, namely Type Ia supernovae
(SNIa), cosmic chronometers (CC), and baryon acoustic
oscillation (BAO) measurements. The analysis is performed
within a Bayesian framework, using the total likelihood
$\mathcal{L}_{\mathrm{tot}}\propto
\exp(-\chi^2_{\mathrm{tot}}/2)$, with
\begin{equation}
\chi^2_{\mathrm{tot}}
=
\chi^2_{\mathrm{SNIa}}
+\chi^2_{\mathrm{CC}}
+\chi^2_{\mathrm{BAO}}.
\label{eq:observational_chi2_total}
\end{equation}

For the supernova sector, we employ the Pantheon+
compilation \cite{Brout:2022vxf}, using the unbinned
sample with $z_{\mathrm{HD}}>0.01$, which contains
1590 light-curve measurements. We use the corrected
apparent magnitudes and the full covariance matrix,
including statistical and systematic uncertainties,
without incorporating the SH0ES distance-ladder
calibration. The theoretical apparent magnitude is
calculated from the luminosity distance
$D_L=(1+z_{\mathrm{HEL}})D_M(z_{\mathrm{HD}})$,
where $z_{\mathrm{HEL}}$ and $z_{\mathrm{HD}}$ denote
the heliocentric and Hubble-diagram redshifts,
respectively. For a spatially flat background,
the transverse comoving distance is
\begin{equation}
D_M(z)=\frac{c}{H_0}
\int_0^z\frac{dz'}{E(z')}.
\label{eq:observational_comoving_distance}
\end{equation}
The absolute magnitude $\mathcal{M}$ is treated as a free
nuisance parameter. The supernova contribution is evaluated
through
\begin{equation}
\chi^2_{\mathrm{SNIa}}
=
\Delta\mathbf{m}^{T}
\mathbf{C}_{\mathrm{SNIa}}^{-1}
\Delta\mathbf{m},
\label{eq:observational_chi2_SNIa}
\end{equation}
where $\Delta\mathbf{m}$ denotes the residual vector
between observed and theoretical magnitudes, and
$\mathbf{C}_{\mathrm{SNIa}}$ is the corresponding
covariance matrix.

For the cosmic-chronometer sector, we employ 31
measurements of $H(z)$, obtained from the differential
ages of passively evolving galaxies
\cite{Jimenez:2001gg,Moresco:2016mzx}.
Their contribution follows from the standard Gaussian
likelihood, comparing the observed values with
$H(z)=H_0E(z)$, assuming uncorrelated measurement
uncertainties.

For the BAO sector, we use the measurements from the
Dark Energy Spectroscopic Instrument Data Release 2
(DESI DR2) \cite{DESI:2025zgx}. The dataset consists
of 13 observables, involving the transverse comoving
distance $D_M(z)/r_d$, the Hubble distance
$D_H(z)/r_d$, and the volume-averaged distance
$D_V(z)/r_d$, where
\begin{equation}
D_H(z)=\frac{c}{H(z)},
\qquad
D_V(z)=\left[zD_M^2(z)D_H(z)\right]^{1/3}.
\label{eq:observational_BAO_distances}
\end{equation}
Here $r_d$ is the sound horizon at the baryon drag
epoch, which we treat as a free parameter.
The BAO likelihood is constructed using the published
covariance matrix, including correlations between
the measured distance ratios. The combined dataset
therefore contains 1590 SNIa, 31 CC, and 13 BAO
observables, corresponding to $N_{\mathrm{data}}=1634$.

The parameter space is explored using the Markov Chain
Monte Carlo (MCMC) method, implemented through the
publicly available \texttt{emcee} package
\cite{Foreman-Mackey:2012any}. We vary
$\Omega_{m0}$, $h=H_0/(100\,\mathrm{km\,s^{-1}\,Mpc^{-1}})$,
$\mathcal{M}$, and $r_d$, together with the
non-conservation parameter $\xi_0$ for Model I
and the additional parameter $\xi_1$ for Model II.
We adopt uniform priors
\begin{align}
0.10&<\Omega_{m0}<0.50, \qquad
0.55<h<0.85, \nonumber\\
-20&<\mathcal{M}<-18, \qquad
130<r_d/\mathrm{Mpc}<165, \nonumber\\
-2.5&<\xi_0<2.5, \qquad
-6<\xi_1<6,
\label{eq:observational_priors}
\end{align}
where the last prior applies only to Model II.
We fix $\Omega_{b0}=0.05$ and
$\Omega_{r0}=9\times10^{-5}$.
For the MCMC analysis, we employ 48 walkers and
40000 steps, assessing convergence through the
integrated autocorrelation times.

The same likelihood and common parameter priors are
used for the reference $\Lambda$CDM and constant-rate
thermogravity models, ensuring a consistent comparison.
Furthermore, the observational constraints are obtained
from the posterior distributions, while the statistical
comparison is based on the minimum chi-square and the
Akaike and Bayesian information criteria
\cite{Liddle:2007fy},
\begin{equation}
\mathrm{AIC}=\chi^2_{\min}+2k,
\qquad
\mathrm{BIC}=\chi^2_{\min}+k\ln N_{\mathrm{data}},
\label{eq:observational_information_criteria}
\end{equation}
where $k$ denotes the number of fitted parameters.
The resulting parameter constraints and model comparison
are presented in the following subsections.

\subsection{Observational results}
\label{subsec:observational_results}

We now present the observational constraints on Models I and II,
obtained from the combined Pantheon+ SNIa, CC, and DESI DR2 BAO
dataset. The parameter estimates correspond to the posterior
medians and the associated $68\%$ credible intervals.

\begin{figure*}[ht!]
\centering
\includegraphics[width=0.85\textwidth]{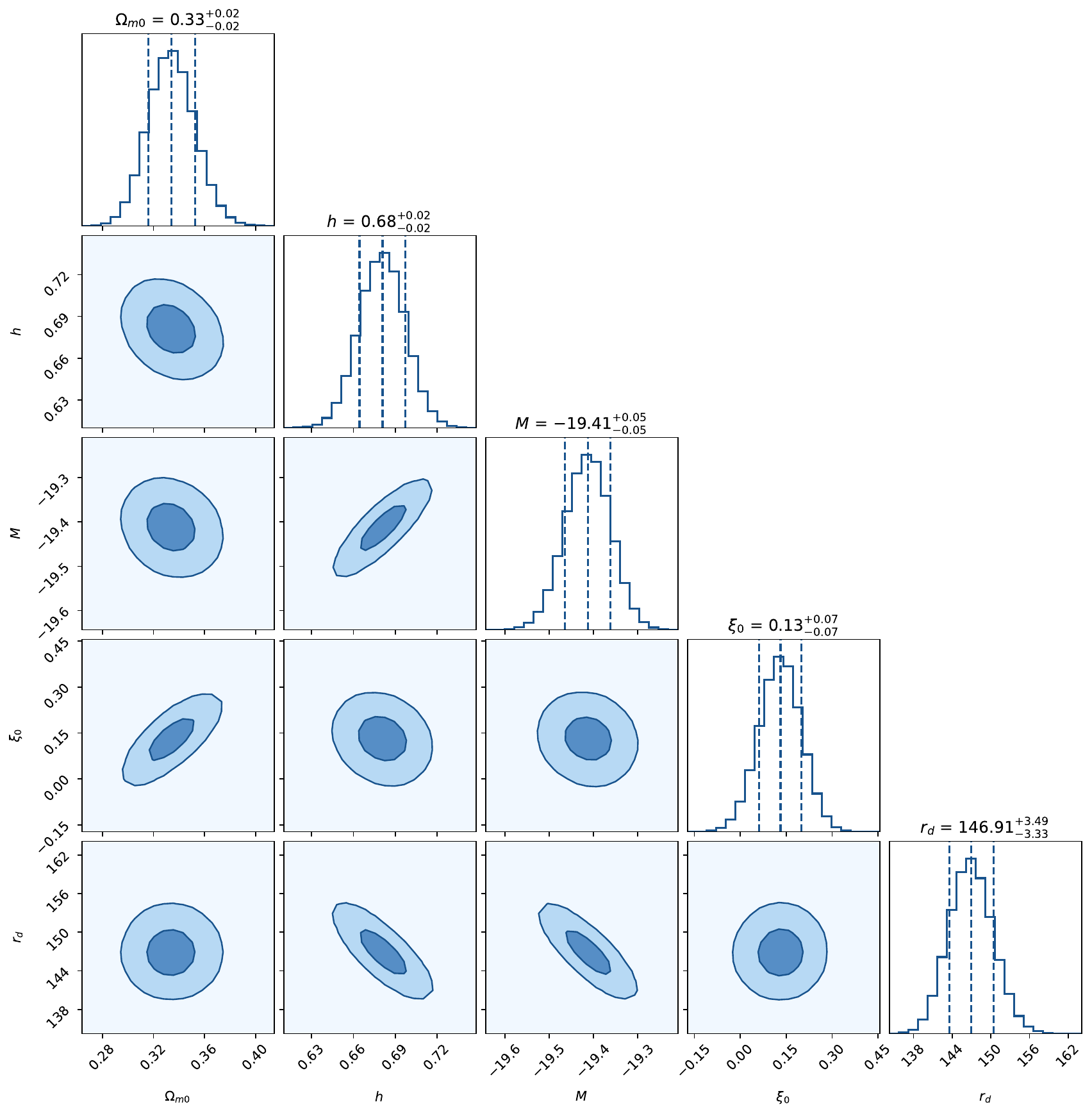}
\caption{\textit{Marginalized posterior distributions
and two-dimensional joint constraints on the parameters
$\Omega_{m0}$, $h$, $\mathcal{M}$, $\xi_0$, and $r_d$
of Model I, defined by Eq.~(\ref{eq:Gamma_model_I}),
obtained from the combined Pantheon+ SNIa,
cosmic-chronometer, and DESI DR2 BAO datasets.
The diagonal panels show the one-dimensional
marginalized distributions, while the off-diagonal
panels display the corresponding joint posterior
contours. The dashed vertical lines indicate the
median values and the $68\%$ credible intervals.
The non-conservation parameter $\xi_0$ exhibits
a mild preference for positive values, although
the standard conservation limit remains compatible
with the data at approximately the $95\%$ credible level.}}
\label{fig:model_I_corner}
\end{figure*}

For Model I, the free parameters are
$\{\Omega_{m0},h,\mathcal{M},\xi_0,r_d\}$.
The resulting constraints are summarized in
Table~\ref{tab:observational_results_model_I},
while the marginalized posterior distributions and the
corresponding two-dimensional joint constraints are displayed in
Fig.~\ref{fig:model_I_corner}.
As we can see, the non-conservation parameter exhibits
a mild preference for positive values, with
$\xi_0=0.131^{+0.068}_{-0.069}$, corresponding to
matter production. Nevertheless, the standard conservation
limit $\xi_0=0$ remains compatible with the data
at approximately the $95\%$ credible level.

\begin{table*}[ht!]
\centering
\caption{Observational constraints on Model I, defined by
Eq.~(\ref{eq:Gamma_model_I}), from the combined Pantheon+
SNIa, cosmic-chronometer, and DESI DR2 BAO datasets.
The parameter estimates correspond to the posterior medians
and the $68\%$ credible intervals, while $\chi^2_{\min}$
denotes the minimum chi-square obtained through numerical
optimization. The number of degrees of freedom is
$\mathrm{dof}=N_{\mathrm{data}}-k=1629$.}
\small
\begin{tabular}{lccccccc}
\hline\hline
Model
& $\Omega_{m0}$
& $h$
& $r_d$ (Mpc)
& $\xi_0$
& $\mathcal{M}$
& $\chi^2_{\min}$
& $\chi^2_{\min}/\mathrm{dof}$ \\
\hline
Model I
& $0.334^{+0.019}_{-0.018}$
& $0.681^{+0.016}_{-0.017}$
& $146.912^{+3.486}_{-3.328}$
& $0.131^{+0.068}_{-0.069}$
& $-19.412^{+0.051}_{-0.052}$
& $1427.339$
& $0.876$ \\
\hline\hline
\end{tabular}
\label{tab:observational_results_model_I}
\end{table*}

An interesting feature of Model I is the positive
correlation between $\xi_0$ and $\Omega_{m0}$.
This can be understood from the modified cold dark
matter evolution,
$\rho_c\propto a^{-3+\xi_0}$.
For a fixed present-day matter density, increasing
$\xi_0$ reduces the cold dark matter abundance
at earlier times. An increase in $\Omega_{m0}$
can partially compensate for this effect in the
expansion history, giving rise to the observed
parameter degeneracy. The correlations involving
$h$, $\mathcal{M}$, and $r_d$ reflect the
combined distance and expansion-rate constraints
of the observational probes.

\begin{figure*}[ht!]
\centering
\includegraphics[width=0.92\textwidth]{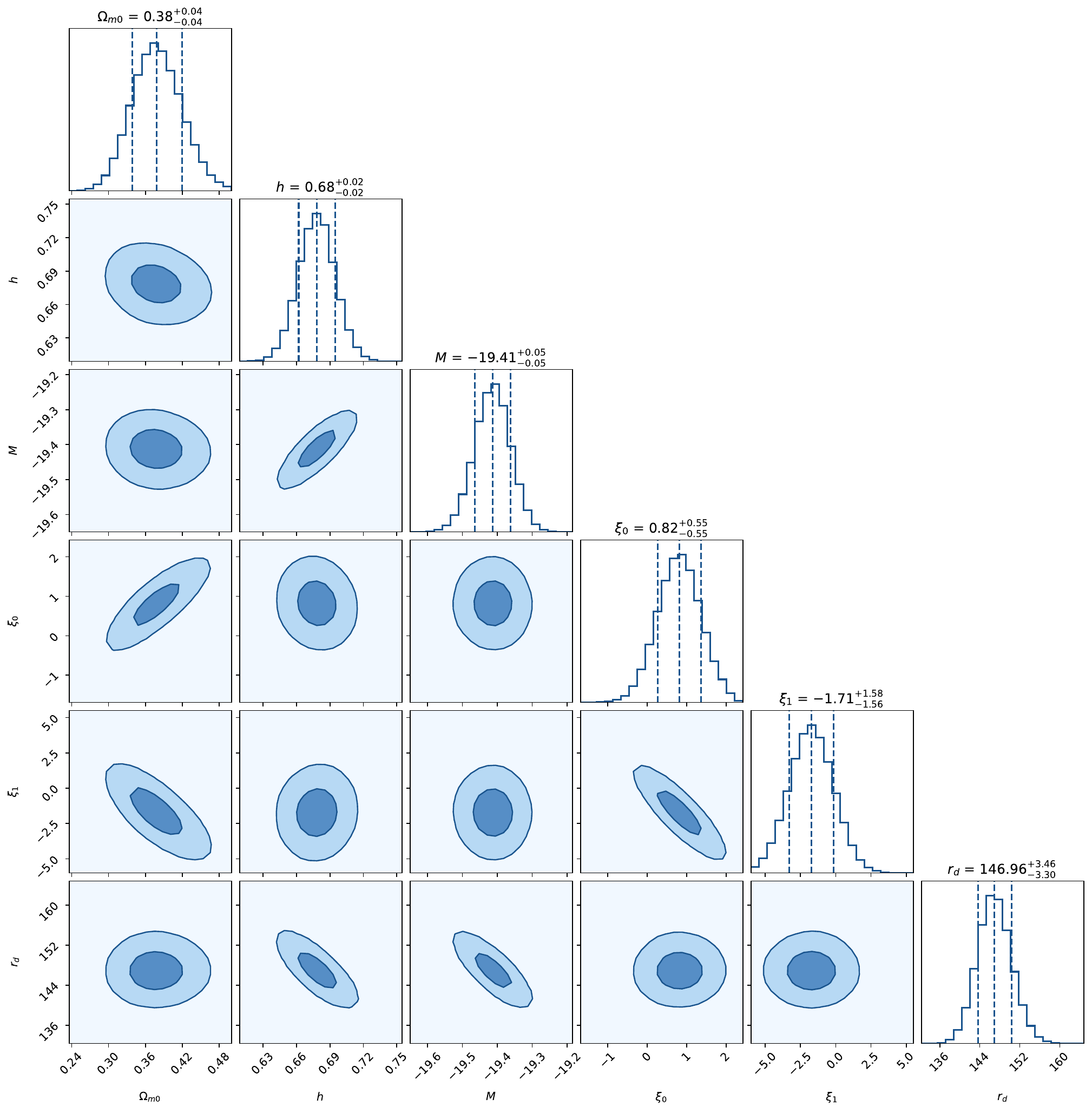}
\caption{\textit{Marginalized posterior distributions
and two-dimensional joint constraints on the parameters
$\Omega_{m0}$, $h$, $\mathcal{M}$, $\xi_0$,
$\xi_1$, and $r_d$ of Model II, defined by
Eq.~(\ref{eq:Gamma_model_II}), obtained from the combined
Pantheon+ SNIa, cosmic-chronometer, and DESI DR2 BAO
datasets. The diagonal panels show the marginalized
distributions, while the off-diagonal panels display
the joint posterior contours. The dashed vertical
lines indicate the median values and the $68\%$
credible intervals. A pronounced negative correlation
between $\xi_0$ and $\xi_1$ is observed, with the
marginalized distributions favoring positive $\xi_0$
and negative $\xi_1$.}}
\label{fig:model_II_corner}
\end{figure*}

We next consider Model II, for which the additional
running parameter $\xi_1$ is included.
The parameter space is therefore
$\{\Omega_{m0},h,\mathcal{M},\xi_0,\xi_1,r_d\}$.
The corresponding marginalized constraints are
presented in Table~\ref{tab:observational_results_model_II},
while the posterior distributions and the
two-dimensional joint constraints are shown in
Fig.~\ref{fig:model_II_corner}.

\begin{table*}[t]
\centering
\caption{Observational constraints on Model II, defined by
Eq.~(\ref{eq:Gamma_model_II}), from the combined Pantheon+
SNIa, cosmic-chronometer, and DESI DR2 BAO datasets.
The parameter estimates correspond to the posterior medians
and the $68\%$ credible intervals, while $\chi^2_{\min}$
denotes the minimum chi-square obtained through numerical
optimization. The number of degrees of freedom is
$\mathrm{dof}=N_{\mathrm{data}}-k=1628$.}
\small
\setlength{\tabcolsep}{4pt}
\begin{tabular}{lcccccccc}
\hline\hline
Model
& $\Omega_{m0}$
& $h$
& $r_d$ (Mpc)
& $\xi_0$
& $\xi_1$
& $\mathcal{M}$
& $\chi^2_{\min}$
& $\chi^2_{\min}/\mathrm{dof}$ \\
\hline
Model II
& $0.378^{+0.042}_{-0.039}$
& $0.678^{+0.016}_{-0.016}$
& $146.957^{+3.463}_{-3.303}$
& $0.818^{+0.549}_{-0.551}$
& $-1.714^{+1.582}_{-1.562}$
& $-19.413^{+0.050}_{-0.052}$
& $1425.652$
& $0.876$ \\
\hline\hline
\end{tabular}
\label{tab:observational_results_model_II}
\end{table*}

The posterior distributions indicate that the
non-conservation parameters are shifted from zero,
with $\xi_0=0.818^{+0.549}_{-0.551}$ and
$\xi_1=-1.714^{+1.582}_{-1.562}$.
Nevertheless, their relatively large uncertainties
and strong correlation do not establish a
statistically compelling departure from standard
matter conservation. In particular, the additional
running parameter $\xi_1$ remains weakly constrained
by the present combination of late-time observations.
The marginalized uncertainty on $\xi_0$ is also
considerably larger than in Model I.

A particularly notable feature of Model II is the
pronounced negative correlation between $\xi_0$
and $\xi_1$. This degeneracy follows from the
functional form of the non-conservation rate,
$\frac{\Gamma(H)}{H_0}
=\xi_0+\xi_1\left(1-\frac{1}{E}\right)$.
Over the redshift range probed by the observations,
an increase in one parameter can be partially
compensated by a decrease in the other, yielding
similar non-conservation histories and hence
similar background observables. Consequently,
the data constrain certain combinations of
$\xi_0$ and $\xi_1$ more effectively than
the individual parameters.

The additional running freedom also broadens
the allowed range of $\Omega_{m0}$, reflecting
its correlations with the non-conservation
parameters. By contrast, the constraints on
$h$, $\mathcal{M}$, and $r_d$ remain very similar
to those obtained for Model I. In particular,
both models yield $h\simeq0.68$ and
$r_d\simeq147\,\mathrm{Mpc}$, with no substantial
difference in their preferred values.

In summary, the combined late-time observations
show a mild preference for positive present-day
non-conservation rates in both models, although
they do not establish a statistically significant
departure from standard cold dark matter
conservation. While Model II allows a more general
evolution of the non-conservation rate, including
possible transitions between matter production
and depletion, the present data do not provide
compelling evidence for this additional freedom.
The implications for the statistical preference
between the models, including the comparison with
the $\Lambda$CDM scenario and constant-rate
thermogravity, are examined in the following
subsection.

\subsection{Model comparison}
\label{subsec:model_comparison}

Having obtained the observational constraints, we now
compare the statistical performance of the two thermogravity
models with that of $\Lambda$CDM and the previously studied
constant-rate thermogravity scenario. All four models are
confronted with the same Pantheon+ SNIa, CC, and DESI DR2
BAO dataset, allowing for a consistent comparison of their
goodness of fit and statistical preference. We employ the
minimum chi-square, $\chi^2_{\min}$, together with the Akaike
and Bayesian information criteria, defined in
Eq.~\eqref{eq:observational_information_criteria}.

The minimum chi-square values are
$\chi^2_{\min}=1430.752$ for $\Lambda$CDM,
$1426.757$ for constant-rate thermogravity,
$1427.339$ for Model I, and $1425.652$ for Model II.
These values are obtained through numerical optimization
and indicate that all models reproduce the observational
data with a comparable goodness of fit. In particular,
the thermogravity scenarios yield smaller
$\chi^2_{\min}$ values than $\Lambda$CDM, with
improvements of approximately $4.00$, $3.41$, and $5.10$
for constant-rate thermogravity, Model I, and Model II,
respectively. Although these improvements remain modest,
they indicate that the introduction of matter
non-conservation can provide a better description of
the combined late-time observations.

To assess whether the improvement in the fit justifies
the additional parameters, we calculate the AIC and BIC.
The results are presented in
Table~\ref{tab:observational_model_comparison},
where the differences are defined relative to
$\Lambda$CDM, namely
$\Delta\mathrm{IC}=\mathrm{IC}_{\mathrm{model}}
-\mathrm{IC}_{\Lambda\mathrm{CDM}}$, with IC denoting
either AIC or BIC. Negative differences therefore
indicate a preference over the standard model.

\begin{table}[t]
\centering
\caption{Information criteria for $\Lambda$CDM,
constant-rate thermogravity (TG), and the two running
thermogravity models, using the combined Pantheon+
SNIa, CC, and DESI DR2 BAO dataset, with
$N_{\mathrm{data}}=1634$.
The differences are defined relative to $\Lambda$CDM
as $\Delta\mathrm{IC}=\mathrm{IC}_{\mathrm{model}}
-\mathrm{IC}_{\Lambda\mathrm{CDM}}$,
where IC denotes AIC or BIC.
Negative values favor the corresponding model over
$\Lambda$CDM.}
\begin{tabular}{lcccc}
\hline\hline
Model & AIC & $\Delta$AIC & BIC & $\Delta$BIC \\
\hline
$\Lambda$CDM
& 1438.752 & 0 & 1460.347 & 0 \\
Constant-rate TG
& 1436.757 & -1.995 & 1463.751 & 3.404 \\
Model I
& 1437.339 & -1.413 & 1464.333 & 3.986 \\
Model II
& 1437.652 & -1.100 & 1470.045 & 9.698 \\
\hline\hline
\end{tabular}
\label{tab:observational_model_comparison}
\end{table}

As shown in Table~\ref{tab:observational_model_comparison},
all three thermogravity scenarios are slightly favored
over $\Lambda$CDM according to the AIC. In particular,
Model I yields $\Delta\mathrm{AIC}\simeq-1.41$,
while the constant-rate thermogravity model gives
$\Delta\mathrm{AIC}\simeq-2.00$.
For Model II, despite its larger improvement in
$\chi^2_{\min}$, the additional running parameter
reduces the AIC preference to
$\Delta\mathrm{AIC}\simeq-1.10$.
These differences are relatively small and do not
establish a strong statistical preference for any
of the thermogravity prescriptions.

The BIC, which assigns a stronger penalty to
additional parameters, continues to favor
$\Lambda$CDM. The corresponding differences are
$\Delta\mathrm{BIC}\simeq3.40$ for constant-rate
thermogravity, $3.99$ for Model I, and $9.70$
for Model II. This behavior reflects the fact
that the improvements in $\chi^2_{\min}$ do not
fully compensate for the increased number of
free parameters under this criterion.
In particular, although Model II provides the
smallest $\chi^2_{\min}$, its additional running
freedom is not statistically preferred over
the simpler thermogravity prescriptions.
Nevertheless, the information-criterion differences
should not be interpreted as an observational
exclusion of these scenarios.

Concerning the magnitude of matter non-conservation,
the posterior constraints show a mild preference
for positive present-day non-conservation rates.
For Model I, the relatively narrow distribution
of $\xi_0$ favors matter production, while
significant departures from standard cold dark
matter dilution remain constrained. A similar
tendency is obtained for constant-rate thermogravity.
In Model II, the pronounced negative correlation
between $\xi_0$ and $\xi_1$ allows a wider range
of non-conservation histories while maintaining
good agreement with observations. Consequently,
although the Pantheon+ analysis provides indications
of nonzero matter non-conservation, the statistical
evidence remains insufficient to establish a
compelling departure from the standard conservation
limit.

In summary, the observational analysis demonstrates
that both running thermogravity models provide
viable descriptions of the late-time cosmological
background, with modest improvements in the
goodness of fit relative to $\Lambda$CDM.
The AIC slightly favors the thermogravity scenarios,
whereas the BIC continues to favor the simpler
standard model, particularly over Model II.
Thus, although the present data show a mild
preference for matter non-conservation, they do
not provide compelling evidence for the additional
running freedom. Future measurements with improved
precision, together with complementary cosmological
probes, may help break the parameter degeneracies
and further distinguish between the different
thermogravity prescriptions.

 \section{Conclusions}
\label{sec:conclusions}

The nature of dark energy and the physical origin of the
late-time accelerated expansion of the Universe remain
among the central open questions of modern cosmology.
While these phenomena are commonly addressed through
additional fields or fluids within general relativity,
or through modifications of the gravitational sector,
the thermodynamic interpretation of gravity offers a
different possibility. In particular, the recently
proposed thermogravity framework extends the usual
thermodynamic derivation of gravitational dynamics by
including work-producing processes, leading to trace-free
field equations and a modified energy-momentum conservation
law. An important consequence is that the cosmological
evolution can be influenced by matter non-conservation
and the associated geometrical contribution, without
introducing a fundamental dark-energy component.

In this work, motivated by the extensive investigation
of running quantities in various cosmological frameworks,
we considered an extension of thermogravity in which
the characteristic non-conservation scale is allowed
to evolve with the cosmological expansion, namely
$L_2=L_2(H)$. We focused on the realization where
non-conservation affects only cold dark matter, while
baryons and radiation retain their standard evolution.
Within this framework, we considered two specific
prescriptions: Model I, with $\Gamma(H)=\xi_0 H$,
and Model II, with
$\Gamma(H)=H_0[\xi_0+\xi_1(1-H_0/H)]$.
We derived the corresponding cosmological equations,
examined their analytical and numerical background
solutions, and investigated the evolution of the
effective geometric dark-energy sector. Furthermore,
we performed a dynamical-system analysis and confronted
both models with late-time cosmological observations
from Type Ia supernovae, cosmic chronometers, and
DESI DR2 baryon acoustic oscillations.

Concerning the background evolution, we found that
the two prescriptions lead to different non-conservation
histories and consequently to distinct cosmological
behavior. Model I admits an exact analytical solution,
with the cold dark matter density evolving as
$\rho_c\propto a^{-3+\xi_0}$, and reproduces the
quadratic running-vacuum form in the dust-only limit.
In contrast, Model II exhibits a suppressed relative
non-conservation rate at early times, allowing the
standard matter dilution law to be recovered
asymptotically. Moreover, its additional running
parameter permits changes in the sign of the
non-conservation rate, corresponding to reversals
in the effective energy exchange between cold dark
matter and the geometrical sector. As a result,
the reconstructed dark-energy contribution can
exhibit quintessence-like or phantom-like behavior,
including smooth crossings of the phantom divide.
Interestingly, such behavior arises despite the
vacuum-like intrinsic equation of state of the
geometrical contribution, and without requiring
a phantom field or superacceleration.

The dynamical-system analysis revealed further
differences between the two models. In particular,
Model I admits a modified matter-scaling critical
point, whereas Model II recovers the standard
matter-dominated critical configurations at the
high-curvature boundary. Both models also possess
a continuous family of de Sitter solutions, which
are transversely attracting when
$\Gamma(H_\star)/H_\star<3$, providing possible
late-time accelerated states. Concerning the
observational analysis, we found that both models
provide a good description of the combined
Pantheon+ SNIa, CC, and DESI DR2 BAO dataset,
with minimum chi-square values slightly smaller
than that of $\Lambda$CDM. In particular, Model I
yields $\xi_0=0.131^{+0.068}_{-0.069}$, indicating
a mild preference for positive matter non-conservation,
while in Model II the additional running freedom
leads to broader constraints and a pronounced negative
correlation between $\xi_0$ and $\xi_1$.
The Akaike information criterion slightly favors
the thermogravity scenarios over $\Lambda$CDM,
whereas the Bayesian information criterion
continues to favor the simpler standard model,
particularly over Model II. Hence, although the
present data show indications of nonzero matter
non-conservation, they do not establish a
statistically compelling departure from the
standard conservation limit, nor do they require
the additional running freedom.

The above results suggest several directions
for further investigation. An important next
step would be to extend the analysis beyond
the homogeneous background and study cosmological
perturbations, the growth of matter fluctuations,
and the corresponding implications for large-scale
structure formation. Such an investigation could
also clarify the differences between thermogravity
and conventional interacting dark-energy models,
whose background evolution can be formally similar.
Furthermore, combining late-time observations with
cosmic microwave background and early-Universe
constraints would allow a more complete assessment
of the different non-conservation prescriptions,
particularly those that remain active at high
redshifts. Finally, deriving the running
non-conservation scale from a microscopic
thermodynamic description, rather than introducing
it phenomenologically, would be an important
step toward understanding its physical origin.
These developments may help establish whether
thermogravity can provide a consistent and
observationally distinguishable description
of cosmological evolution.

 \begin{acknowledgments}  The author acknowledges the contributions of the LISA
CosWG  and  COST
Actions
 CA21106 ``COSMIC WISPers
in the Dark Universe: Theory, astrophysics and experiments'',  CA21136
``Addressing observational tensions in cosmology with
  systematics and fundamental physics (CosmoVerse)'',    CA23130
``Bridging high and low energies in
search of quantum gravity (BridgeQG)'', and CA24101 ``Testing Fundamental
Physics with Seismology''.
 \end{acknowledgments}
 
 
\end{document}